\documentclass[aps,prl,reprint,superscriptaddress,twocolumn,showkeys,amsmath,amssymb,nofootinbib]{revtex4-1}
\usepackage[utf8]{inputenc}
\usepackage{bm}
\usepackage{graphicx}
\usepackage{amsthm}
\usepackage[english]{babel}
\usepackage{microtype}
\usepackage{mathtools}
\usepackage{dsfont}
\usepackage{mathrsfs}
\usepackage{braket}
\usepackage{hyperref}
\usepackage[shortlabels]{enumitem}
\usepackage[dvipsnames]{xcolor}
\usepackage{stmaryrd}
\usepackage{afterpage}
\usepackage{marvosym}

\hypersetup{
  colorlinks   = true, 
  urlcolor     = Blue, 
  linkcolor    = BrickRed, 
  citecolor   = PineGreen 
}

\newcommand{\Zstab}{S^{Z}} %
\newcommand{\Xstab}{S^{X}} %
\newcommand{\Xlog}{\overline{X}} %
\newcommand{\Zlog}{\overline{Z}} %
\newcommand{\hcl}{q} %
\newcommand{\Mtot}{\mathcal{M}} %
\newcommand{\Stot}{\mathcal{S}} %
\newcommand{\Ttot}{\mathcal{T}} %
\newcommand{\Rtot}{\mathcal{R}} %

\makeatletter
\def\@fnsymbol#1{\ensuremath{\ifcase#1\or \!a\or \hspace*{-0.1em}b\or * \or \dagger\or \ddagger\or
   \mathsection\or \mathparagraph\or \|\or  \dagger\dagger
   \or \ddagger\ddagger \else\@ctrerr\fi}}
\makeatother

\DeclareMathOperator{\sech}{sech}
\DeclareMathOperator{\sgn}{sgn}
\DeclareMathOperator{\Tr}{Tr}

\pdfpageheight\paperheight
\pdfpagewidth\paperwidth

\def\tcm{T.C.M. Group, Cavendish Laboratory, University of Cambridge, J.J. Thomson Avenue, Cambridge, CB3 0HE, UK}
\def\DAMTP{DAMTP, University of Cambridge, Wilberforce Road, Cambridge, CB3 0WA, UK}

\begin{document}

\title{Coherent error threshold for surface codes from Majorana delocalization}

\author{Florian Venn}
\thanks{These authors contributed equally.}
\affiliation{\DAMTP}

\author{Jan Behrends}
\thanks{These authors contributed equally.}
\affiliation{\tcm}

\author{Benjamin B\'eri}
\email[]{bfb26@cam.ac.uk}
\affiliation{\DAMTP}
\affiliation{\tcm}

\begin{abstract}
Statistical mechanics mappings provide key insights on quantum error correction. 
However, existing mappings assume incoherent noise, thus ignoring coherent errors due to, e.g., spurious gate rotations. 
We map the surface code with coherent errors, taken as $X$- or $Z$-rotations (replacing bit or phase flips), to a two-dimensional (2D) Ising model with complex couplings, and further to a 2D Majorana scattering network. 
Our mappings reveal both commonalities and qualitative differences in correcting coherent and incoherent errors. 
For both, the error-correcting phase maps, as we explicitly show by linking 2D networks to 1D fermions, to a $\mathbb{Z}_2$-nontrivial 2D insulator.
However, beyond a rotation angle $\phi_\text{th}$, instead of a $\mathbb{Z}_2$-trivial insulator as for incoherent errors, coherent errors map to a Majorana metal. 
This $\phi_\text{th}$ is the theoretically achievable storage threshold.
We numerically find $\phi_\text{th}\approx0.14\pi$.
The corresponding bit-flip rate $\sin^2(\phi_\text{th})\approx 0.18$ exceeds the known incoherent threshold $p_\text{th}\approx0.11$.
\end{abstract}

\maketitle

A major milestone towards building scalable quantum computers is quantum error correction (QEC)~\cite{CalShor96,Steane96,TerhalRMP}. 
Surface codes are among the most promising candidates for this~\cite{kitaev1997quantum,kitaev2003fault,BravyiKitaev_SC,dennis2002topological,Fowler12}.
Their layout informs the design of state-of-the-art many-qubit devices~\cite{Arute_QS}, where most recent developments include proof-of-principle demonstrations of surface-code QEC on small systems~\cite{krinner2022realizing,Google_SC}.

Key insights on the phenomenology and fundamental performance limits of QEC codes come from mappings to statistical mechanics models~\cite{dennis2002topological,wang2003confinement,Katzgraber09,Bombin12,Kubica18,chubb2021statistical}. 
For the surface code, assuming ideal measurements and either bit-flip $X$ or phase-flip $Z$ errors occurring with probability $p$, this is the two-dimensional (2D) random-bond Ising model (RBIM)~\cite{dennis2002topological,wang2003confinement}. 
The ordered and disordered RBIM phases map, respectively, to regimes where QEC succeeds and fails for large system size $L$, while the phase transition marks the theoretical maximum rate \mbox{$p_\text{th}\approx 0.11$~\cite{Honecker01,MerzChalker02,Suchara14}} of errors that can be corrected. (Tailoring the code for such biased noise may achieve higher thresholds~\cite{Tuckett:2020ec}.)

The RBIM mapping assumes incoherent errors. %
Coherent errors can, however, also 
arise, e.g., from unintended or imperfect gate rotations~\cite{Kueng16,Wallman15,Wallman16,Debroy18,Hashim21,greenbaum2017modeling,Beale18,HuangDohertyFlammia19,bravyi2018correcting,iverson2020coherence,Gottesman19,Florian20}.
While results are favorable on their mitigation~\cite{Wallman16,Debroy18,Hashim21} or correction at fixed $L$~\cite{Beale18,HuangDohertyFlammia19},
a key question for surface codes is how coherent errors' interference~\cite{iverson2020coherence,Gottesman19} impacts the scaling with $L$. 
Numerical results for either $\exp(i\phi X)$ or $\exp(i\phi Z)$ acting on each qubit suggest that surface code QEC may succeed for $\phi<\phi_\text{c}\approx0.1\pi$~\cite{bravyi2018correcting}. 
While $\phi_\text{c}$ is decoder specific, $\sin^2(\phi_\text{c})\lesssim p_\text{th}$ suggests that assuming bit flips with $p=\sin^2(\phi)$ (``Pauli twirling") may work in practice. 
However, fundamental questions remain: What is the theory, replacing the RBIM, for the QEC phases?
How does the phenomenology of these phases differ from the incoherent case?
What is the maximum achievable threshold $\phi_\text{th}$? 

\begin{figure}[t]
 \includegraphics[scale=1.4]{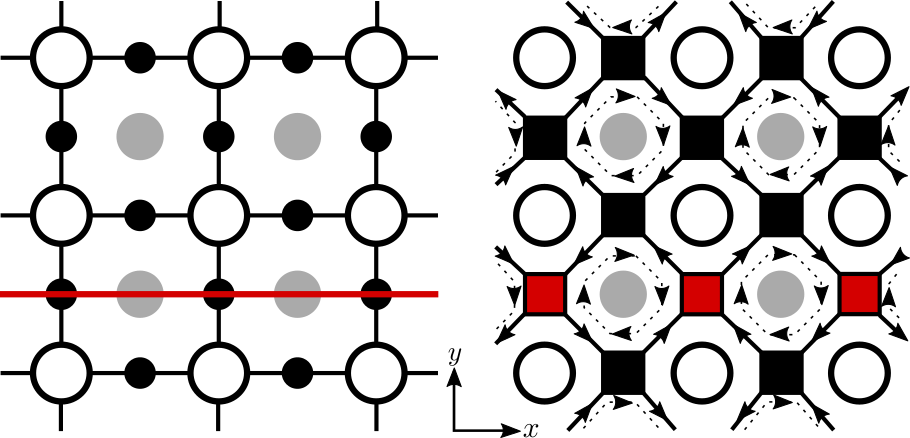}
 \caption{Left panel: bulk patch of a surface code.  
 Black dots are qubits, white and gray disks, respectively, show stabilizers $\Xstab_v$ and $\Zstab_w$. 
 The red line marks the logical $\overline{X}$'s path (for suitable boundary conditions). 
 In the RBIM, the $\Xstab_v$ become spins interacting with their nearest neighbors through couplings set by the errors. 
 Right panel: In the network model, the Ising bonds become junctions scattering incoming into outgoing Majorana modes. 
 Solid and dashed lines show, respectively, the modes' propagation direction for coherent and incoherent errors.  
 In the transfer matrix, the junctions are quantum gates acting on pairs of Majorana operators. 
  }\label{fig:mappings}
\end{figure}

Here we introduce an RBIM that provides such a theory.
Unlike probabilities of incoherent errors, quantum amplitudes now yield complex Boltzmann weights. 
Yet, the problem has two useful and interrelated~\cite{MerzChalker02,Fulga11,Fulga12,Jian22} formulations, each encompassing both incoherent and coherent errors (cf. Fig.~\ref{fig:mappings}): 
a 2D Majorana network, and a 1D fermion Hamiltonian, both arising from the transfer matrix---a non-unitary quantum circuit akin to those of current interest in quantum dynamics~\cite{Fisher22}.

We find that, upon increasing $\phi$, the network undergoes an insulator-metal transition. 
This is qualitatively distinct from the incoherent case whose network, upon increasing $p$, has an insulator-insulator transition~\cite{SenthilFisher2000,ReadLudwig2000,Gruzberg01,Chalker01,MerzChalkerDC,MerzChalker02,EversMirlin08}. 
A key shared feature we find is that both the coherent and the small-$p$ incoherent insulators are $\mathbb{Z}_2$-nontrivial: they correspond to topological 1D fermion phases~\cite{kitaev2001unpaired}. 
We use this to show that correcting coherent errors can succeed for $\phi<\phi_\text{th}$, with $\phi_\text{th}$ the value at the insulator-metal transition. 
We numerically find $\phi_\text{th}\approx 0.14\pi$ for the geometry in Fig.~\ref{fig:mappings}. 
Remarkably, the Pauli-twirled probability $\sin^2(\phi_\text{th})\approx 0.18$ exceeds $p_\text{th}\approx 0.11$ by 64\%.

\textit{QEC ingredients:}
Surface codes are stabilizer codes~\cite{kitaev1997quantum,kitaev2003fault,BravyiKitaev_SC,dennis2002topological,Fowler12,Gottesman96,Calderbank97,Gottesman97,TerhalRMP}. 
They encode logical qubits in the common $+1$ eigenspace of stabilizers $\Xstab_v=\prod_{j\in v}X_j$ at vertices $v$ and $\Zstab_w=\prod_{j\in w}Z_j$ at plaquettes $w$ of the lattice, where we multiply Pauli $X$ and $Z$ operators on qubits around $v$ or $w$ (cf. Fig.~\ref{fig:mappings}). 
The logical Pauli $\Xlog_\gamma=\prod_{j\in \gamma} X_j$ and $\Zlog_{\gamma'}=\prod_{j\in \gamma'} Z_j$ run along noncontractible paths $\gamma^{(\prime)}$ such that they commute with all the stabilizers. 
The logical operators are not unique; their path can be deformed via multiplying by stabilizers. 
We focus on codes with a single logical qubit. 
We denote the logical operators along their shortest possible paths by $\Xlog$ and $\Zlog$, and take $L$ to be $\Xlog$'s path length. 

During the operation of the code, the constituent qubits can suffer errors. 
Here we focus on $X$ errors, namely on the coherent $U=\prod_j e^{i\phi X_j}$, which we shall compare with incoherent bit flips $X_j$ occurring with probability $p=\sin^2(\phi)$. 
($Z$ errors work similarly, upon $\Xstab_v, \Xlog\leftrightarrow\Zstab_w, \Zlog$ below.)
Expanding $U$ we find a superposition of $X$-strings applied to the initial logical state $\ket{\psi}$. 
For a given string, the endpoints are where $\Zstab_w$ have eigenvalue $s_w=-1$. 
Starting from a string $C_s$ consistent with the syndrome $s=\{s_w\}$, we can get all other such strings from multiplying by the $\Xstab_v$ and/or $\overline{X}$. 
The former leaves $C_s\!\ket{\psi}$ unchanged, the latter takes it to $(C_s\overline{X})\!\ket{\psi}$. 
To correct errors, one measures $s$ via the $\Zstab_w$, and then applies either $C_s$ or $C_s\overline{X}$ to return to the logical subspace~\cite{TerhalRMP}. 
In practice, this choice is made by a decoder. 
But if aiming for the theoretical optimum, one maximizes~\cite{Florian20}
$P_{\hcl,s}=|\langle{\psi}|C_s\overline{X}^\hcl U|\psi\rangle|^2$ over $\hcl=0,1$.
Henceforth, we take $\ket{\psi}$ to be a $\overline{Z}$ eigenstate; then, from $ \overline{X}^\hcl\!\ket{\psi}$ being orthogonal, $P_{\hcl,s}$ are probabilities. 
The considerations in the incoherent case are similar, but instead of a superposition, we define $P_{\hcl,s}$ for a probabilistic ensemble of strings~\cite{dennis2002topological,TerhalRMP}.
In both cases, the probability of syndrome $s$ is $P_s=\sum_\hcl P_{\hcl,s}$.

The feasibility of QEC hinges on $P_{0,s}$ and $P_{1,s}$ being sufficiently distinct. 
We measure this via
\begin{equation}\label{eq:Delta}
\Delta=\sum_{s} P_{s}\min_{\hcl}\frac{P_{\hcl,s}}{P_{s}}=\sum_s \min_{\hcl}P_{\hcl,s}.
\end{equation}
Besides its meaning for $\overline{Z}$ eigenstates  $\ket{\psi}$, due to $P_{\hcl,s}=\frac{3}{2}P_s\delta^{(\hcl,s)}$ with $\delta^{(\hcl,s)}$ the Bloch-sphere-averaged infidelity between pre- and post-QEC states~\cite{Florian20}, $\Delta$ also sets the minimal average infidelity.
For incoherent errors, $\Delta$ is the logical error probability for maximum likelihood decoding~\cite{Suchara14}. 
The error correcting phase is defined by $\Delta$ decreasing to zero exponentially with $L$. 
$\Delta$ decaying to zero also marks the decoherence of logical noise~\cite{Florian20,SM}.

\textit{From surface codes to Ising models:} To map our problem to an Ising model, we adapt the derivation of Ref.~\onlinecite{dennis2002topological} to the coherent case~\cite{SM}. 
In terms of the expansion of $U=\prod_j (\openone \cos \phi + i X_j \sin \phi)$ in $X$-strings, $\langle{\psi}|C_s \overline{X}^\hcl U|\psi\rangle$ is
the sum of coefficients of $C_s \overline{X}^\hcl\prod_v(\Xstab_v)^{r_v}$ for various configurations of $r_v\in\{0,1\}$. 
(Other terms contribute to $U\ket{\psi}$ with states orthogonal to $C_s \overline{X}^\hcl|\psi\rangle$.)
In an $N$-qubit system, $C_s \overline{X}^\hcl$ %
(i.e., all $r_v=0$)
has coefficient
\begin{equation}\label{eq:coeff}
(i\sin \phi)^n(\cos \phi)^{N-n} = \mathcal{N}\prod_j e^{-\eta^{(\hcl,s)}_j J}
\end{equation}
with $n$ the Pauli weight of $C_s \overline{X}^\hcl$, $e^J=\sqrt{i\tan\phi}$ and  $\mathcal{N}=\prod_j\sqrt{i\sin\phi\cos\phi}$; the signs are $\eta^{(\hcl,s)}_j=-1$ if $C_s \overline{X}^\hcl$ includes $X_j$, $\eta^{(\hcl,s)}_j=1$ otherwise. 
We define Ising spins as $\sigma_v=-(-1)^{r_v}$. 
Since $\sigma_v\sigma_{v'}=1$ for this term, and since each qubit is uniquely specified by nearest neighbor (n.n.) $\Xstab_v$ (cf.~Fig.~\ref{fig:mappings}; we use boundary conditions that also respect this~\cite{SM}), Eq.~\eqref{eq:coeff} equals
$\mathcal{N}\prod_{v,v'\text{n.n.}} e^{-J \eta^{(\hcl,s)}_{vv'} \sigma_v\sigma_{v'}}$. 
(We relabeled $\eta^{(\hcl,s)}_j\mapsto \eta^{(\hcl,s)}_{vv'}$.)
This holds also for other $r_v$ configurations because, by $X_j^2=\openone$, a factor $X_j$ comes from $\prod_v(\Xstab_v)^{r_v}$ only when $r_v=1$ precisely  for one $v$ adjacent to $j$ (thus $\sigma_v\sigma_{v'}=-1$), and this introduces $X_j$ to $C_s\overline{X}^\hcl$ when $\eta^{(\hcl,s)}_j=1$ and removes $X_j$ when $\eta^{(\hcl,s)}_j=-1$. 
Hence, $\langle{\psi}|C_s\overline{X}^\hcl U|\psi\rangle=\mathcal{N}\mathcal{Z}_{\hcl,s}$ with the Ising partition function
\begin{equation}\label{eq:Ising}
\mathcal{Z}_{\hcl,s}=\sum_{\{\sigma_v\}} e^{-J \sum_{v,v'\text{n.n.}}\eta^{(\hcl,s)}_{vv'} \sigma_v\sigma_{v'}},\  e^J\!=\!\sqrt{i\tan\phi}.
\end{equation}
In the incoherent case, 
instead of terms in $U$ we enumerate the probabilities of $C_s \overline{X}^\hcl\prod_v(\Xstab_v)^{r_v}$ $X$-strings; 
hence $i\sin\phi\mapsto p$, $\cos\phi\mapsto 1-p$ above and $P_{\hcl,s}=\mathcal{N}\mathcal{Z}_{\hcl,s}$.
When sampling $P_{s}$ by sampling $C_s$, this is the RBIM on the Nishimori line~\cite{Nishimori81,dennis2002topological}. 
Eq.~\eqref{eq:Ising} is our first key result.

\textit{2D networks and 1D Hamiltonians:} While $\mathcal{Z}_{\hcl,s}$, being complex, might elude a direct statistical physics interpretation, valuable insights arise upon expressing it via the transfer matrix $\hat{\Mtot}_{\hcl,s}$.
It will be useful to construct $\hat{\Mtot}_{\hcl,s}$ along $\overline{X}$'s  path, i.e., the $x$ axis in Fig.~\ref{fig:mappings}. 
The steps being standard~\cite{sachdev_2011,fradkin2013field,SM}, we just state the result:  $\mathcal{Z}_{\hcl,s}=\!\bra{\alpha_L}\! \hat{\Mtot}_{\hcl,s} \!\ket{\alpha_0}$, with $\!\ket{\alpha_r}$ encoding boundary conditions at $x=r$, where $\hat{\Mtot}_{\hcl,s}$ is a quantum circuit
\begin{equation}\label{eq:M}
\hat{\Mtot}_{\hcl,s} = \hat{V}^{(L)}_{\hcl,s}\hat{H}^{(L)}_{\hcl,s}\ldots\hat{V}^{(2)}_{\hcl,s}\hat{H}^{(2)}_{\hcl,s} \hat{V}^{(1)}_{\hcl,s}.
\end{equation}
For system size $M$ along $y$,  $\hat{V}^{(k)}_{\hcl,s}=\otimes_{j=1}^M \hat{v}^{(j,k)}_{\hcl,s}$ and $\hat{H}^{(k)}_{\hcl,s}=\otimes_{j=1}^M A^{(j,k)}_{\hcl,s}\hat{h}^{(j,k)}_{\hcl,s}$ in terms of gates $\hat{v}^{(j,k)}_{\hcl,s}$ and $\hat{h}^{(j,k)}_{\hcl,s}$ arising from the $(j,k)^\text{th}$ vertical and horizontal bond  of the Ising model, respectively [here $\cramped{A^{(j,k)}_{\hcl,s}=\sqrt{2\sinh(2J\eta\smash{_{jk}}^{(\hcl,s)})}}$]. %
Upon a Jordan-Wigner transformation using $2M$ Majorana fermions $\hat{\gamma}_j=\hat{\gamma}_j^\dagger$, $\{\hat{\gamma}_i,\hat{\gamma}_j\}=2\delta_{ij}$, we have 
$\hat{v}_{\hcl,s}^{(j,k)}=e^{-i\kappa_{\hcl,s}^{j,k}\hat{\gamma}_{2j}\hat{\gamma}_{2j+1}}$ ($j<M$) and $\hat{v}_{\hcl,s}^{(M,k)}=e^{i\hat{P}\kappa_{\hcl,s}^{M,k,}\hat{\gamma}_{2M}\hat{\gamma}_{1}}$ with $\hat{P}=(-i)^{M}\hat{\gamma}_{1}\hat{\gamma}_{2}\ldots\hat{\gamma}_{2M}$ the conserved 
fermion parity, and $\hat{h}_{\hcl,s}^{(j,k)}=e^{-i\tilde{\kappa}_{\hcl,s}^{j,k}\hat{\gamma}_{2j-1}\hat{\gamma}_{2j}}$.
(We take $y\equiv y+M$, i.e., a cylinder; this can be argued to capture all key features~\cite{SM}.)
Here $\kappa_{\hcl,s}^{j,k}=J\eta_{jk}^{(\hcl,s)}$\!\!, and $\tilde{\kappa}_{\hcl,s}^{j,k}=-\frac{1}{2}\ln\tanh(J\eta_{jk}^{(\hcl,s)})$.

The (nonunitary) gates $\hat{v}_{\hcl,s}^{(j,k)}$ ($j<M$) and $\hat{h}_{\hcl,s}^{(j,k)}$ act on n.n. fermions (cf. Fig.~\ref{fig:mappings}): they are quadratic in $\hat{\gamma}_j$. 
The same holds for $\hat{v}_{\hcl,s}^{(M,k)}$, and hence also for $\hat{\Mtot}_{\hcl,s}$, for each of $P=\pm1$.
This has two key consequences. 
Firstly, we can write $\hat{\Mtot}_{\hcl,s}\hat{\Mtot}_{\hcl,s}^\dagger=e^{-L\hat{\mathcal{H}}_{\hcl,s}}$ as a thermal density matrix, at inverse temperature $L$, with 1D Hamiltonian $\hat{\mathcal{H}}_{\hcl,s}$ that is free-fermionic for each of $P=\pm 1$~\cite{cylinderfn}.
Taking $L\varepsilon^{(1)}_{\hcl,s}\gg1$  [with $\varepsilon^{(1)}_{\hcl,s}$ the smallest excitation energy] yields the ground state $|\varphi_0\rangle$ which, by the singular value decomposition of $\hat{\Mtot}_{\hcl,s}$, is the steady state of the circuit Eq.~\eqref{eq:M}~\cite{SM}.
We shall link the topology of $|\varphi_0\rangle$ to error correction.

Secondly, the $2M\times 2M$ matrix $\Mtot_{\hcl,s}$, implementing $\hat{\Mtot}_{\hcl,s}\hat{\gamma}_j\hat{\Mtot}_{\hcl,s}^{-1}=(\Mtot_{\hcl,s})_{lj}\hat{\gamma}_l$, arises from a network of 2$\times$2 matrices $v_{\hcl,s}^{(j,k)}=e^{2Y\kappa_{\hcl,s}^{j,k}}$ $(j<M$), $v_{\hcl,s}^{(M,k)}=e^{-2P Y\kappa_{\hcl,s}^{M,k}}$ with $P=\pm 1$, and $h_{\hcl,s}^{(j,k)}=e^{2Y\tilde{\kappa}_{\hcl,s}^{j,k}}$ (here $Y=iXZ$). 
In the incoherent case, as $J$ is real, these are pseudounitary~\cite{MerzChalker02}: $t^\dagger Z t= Z$, with $t=v_{\hcl,s}^{(j,k)}$ or $t=h_{\hcl,s}^{(j,k)}$. 
One can thus
interpret them as acting on a pair $c=\left(\begin{smallmatrix}
c_n\\
c_{n+1}
\end{smallmatrix}\right)$ of counterpropagating modes, conserving their current $c^\dagger Z c$.
The RBIM thus maps to quantum transport~\cite{Cho97,SenthilFisher2000,ReadLudwig2000,Gruzberg01,Chalker01,MerzChalkerDC,MerzChalker02,EversMirlin08}: 
we get a Chalker-Coddington network model~\cite{chalker1988percolation}, with $h_{\hcl,s}^{(j,k)}$ and $v_{\hcl,s}^{(j,k)}$ as junction transfer matrices (cf. Fig.~\ref{fig:mappings}). 
The junction scattering matrices, mapping incoming to outgoing amplitudes $\left(\begin{smallmatrix}
\text{in}_n\\
\text{in}_{n+1}
\end{smallmatrix}\right)$
and 
$\left(\begin{smallmatrix}
\text{out}_n\\
\text{out}_{n+1}
\end{smallmatrix}\right)$,
in suitable phase conventions, are 
 $S_h=\left(\begin{smallmatrix}
a & b \\
b & -a
\end{smallmatrix}\right)$
 and
 $S_v=\left(\begin{smallmatrix}
-b & a\\
a & b
\end{smallmatrix}\right)$ with $a=\sech(2\kappa_{\hcl,s}^{j,k})$, $b=\tanh(2\kappa_{\hcl,s}^{j,k})$~\cite{Chalker01,MerzChalker02,EversMirlin08}. 

We find a different network in the coherent case. 
From $\tilde{\kappa}_{\hcl,s}^{(j,k)}=i\phi-\frac{1}{2}\ln[-\eta_{jk}^{(\hcl,s)}]$, the $h_{\hcl,s}^{(j,k)}$ are unitary. 
This conserves $c^\dagger c$; this is the current if $c$ has copropagating modes.
Furthermore, now $Xv_{\hcl,s}^{(j,k)}$ is pseudo-unitary: 
If $c$'s modes counterpropagate, $v_{\hcl,s}^{(j,k)}$ swaps their direction. 
Equivalently, $v_{\hcl,s}^{(j,k)}$ has a pair of vertically copropagating modes.
In the coherent case, thus, both $h_{\hcl,s}^{(j,k)}$ and $v_{\hcl,s}^{(j,k)}$ have copropagating modes, moving horizontally and vertically, respectively (cf. Fig.~\ref{fig:mappings}).
In a suitable phase convention, the scattering matrices are 
$S_\shortrightarrow=-\eta_{jk}^{(\hcl,s)}\left(\begin{smallmatrix}
\cos(2\phi) & \sin(2\phi)\\
-\sin(2\phi) & \cos(2\phi)
\end{smallmatrix}\right)$, 
$
S_\shortleftarrow = S_\downarrow = S_\shortrightarrow^\dagger$, and $S_\uparrow =-S_\downarrow$,
with arrows for
the %
transmission direction, and %
$\left(\begin{smallmatrix}
\text{out}_n\\
\text{out}_{n+1}
\end{smallmatrix}\right)=S_\alpha \left(\begin{smallmatrix}
\text{in}_n\\
\text{in}_{n+1}
\end{smallmatrix}\right)$ 
with $n$ increasing along $y$ for $\alpha=\shortleftarrow,\shortrightarrow$ 
and along $x$ for $\alpha=\uparrow,\downarrow$.

Their scattering matrices being real, both networks belong to Altland and Zirnbauer's symmetry class D~\cite{AZ97}, with links interpretable as Majorana modes.
We will also consider the networks together with their time-reversed copies. 
This gives time-reversal invariant networks in class DIII.
Viewed as such, the incoherent and coherent cases correspond to, respectively, the spin-conserving and spin-flip limits of the class DIII networks of Ref.~\onlinecite{FulgaDIII}, albeit with $\eta_{jk}^{(\hcl,s)}$ creating a different form of disorder. 

This disorder has the same net effect in the incoherent and coherent cases:
$\eta_{jk}^{(\hcl,s)}=-1$ adds a ``vortex" at each of the adjacent $\Zstab_w$ (cf. Fig.~\ref{fig:mappings}): a mode encircling either of these picks up an extra $\pi$ phase.
With several $\eta_{jk}^{(\hcl,s)}=-1$, vortices appear where $s_w=-1$. %
Vortices are thus the network form of the syndrome $s$.

\textit{Network model phases:} 
In the incoherent case, the network is known to have two insulating (i.e., localized) phases with a transition at $p_\text{th}\approx 0.11$~\cite{ReadLudwig2000,Gruzberg01,Chalker01,MerzChalker02,EversMirlin08}.
Being insulators, the average conductivity $g=\frac{L}{M}\langle\text{Tr}\left(\Ttot^\dagger \Ttot\right)\rangle$ satisfies $g\propto e^{-2L/\xi}$ for $L\gg \xi$;
here $\langle\ldots\rangle$ denotes disorder average, $\Ttot$ is the transmission matrix in the transmission-reflection grading of the total scattering matrix $\Stot=\left(\begin{smallmatrix}
\Rtot & \Ttot' \\
\Ttot & \Rtot'
\end{smallmatrix}\right)$~\cite{BeenRMT}, and $\xi$ is the localization length. 
The two insulators are topologically distinct: for $\mathcal{Q}=\sgn\det(\Rtot_\text{pbc}'\Rtot_\text{apbc}')$, equal to the $\mathbb{Z}_2$ invariant of the doubled class DIII system~\cite{Fulga12,FulgaDIII,SM}, and where (a)pbc denotes (anti)periodic boundary conditions in the transverse direction, we have $\mathcal{Q}=\sgn(p-p_\text{th})$~\cite{MerzChalker02,FulgaDIII}.

In the coherent case, we focus on $0\leq\phi\leq\pi/4$; this includes all inequivalent $\phi$ values~\cite{pi4fn}.
In the clean case (i.e., all $\eta_{jk}^{(\hcl,s)}=1$), any $\phi<\pi/4$ gives a $\mathcal{Q}=-1$ insulator~\cite{FulgaDIII}. 
With disorder, we now argue that the system remains insulating for $0<\phi\ll1$. 
We use that if vortices typically appear in dilute configurations of nearby pairs, then, by the splitting of vortex-induced zero modes, a nearly decoupled network (i.e., with $\phi,p\ll1$ nodes) is an insulator~\cite{Chalker01}.
The typical vortex configurations for $\phi\ll1$ are similar to the $p\ll1$ incoherent case: 
there, for a configuration $s$ with $\omega\ll LM$ adjacent vortex pairs (avp), hence low avp density $n_\omega=\omega/LM$, we have $P_s\propto p^\omega(1-p)^{LM-\omega}$. 
There are $\sim\binom{LM}{\omega}$ such $s$ with similar $P_s$. 
Thus $\omega$ \mbox{has roughly binomial} distribution. Hence, $\langle n_\omega\rangle\propto p$ with variance $\sigma_{n_\omega}^2\propto p/LM$ suppressed for large $LM$. 
Among the $s$ at $n_\omega\approx p$, those with a nonzero density of farther separated vortex pairs (from avp chains) give just a  $\propto e^{-cpLM}$ ($c>0$) fraction of configurations. 
In the coherent case, $|\sum_j a_j|\leq \sum_j|a_j|$ gives $|\mathcal{Z}_{\hcl,s}|\leq \mathcal{Z}_{\hcl,s}(i\sin\phi\mapsto \sin\phi)$; the latter is the incoherent $\mathcal{Z}_{\hcl,s}$ with $p\mapsto \sin\phi$, $1-p\mapsto \cos\phi$ (not the Pauli twirl).
Hence $P_{s}\!\lesssim(\sin^{2}\phi)^\omega(\cos^2\!\phi)^{(LM-\omega)}$ for $\phi\ll1$. 
From here, the previous logic applies: 
vortices typically appear in nearby pairs. Hence, the system insulates for $0<\phi\ll1$.
Since $\mathcal{Q}$ cannot change without delocalization~\cite{SM}, 
and since $\mathcal{Q}=-1$ for $\phi=0$ (a clean system),
this small-$\phi$ insulator has $\mathcal{Q}=-1$.
As we shall show, this implies that QEC succeeds up to a nonzero $\phi_\text{th}$. 

\begin{figure}[t]
 \includegraphics[scale=1]{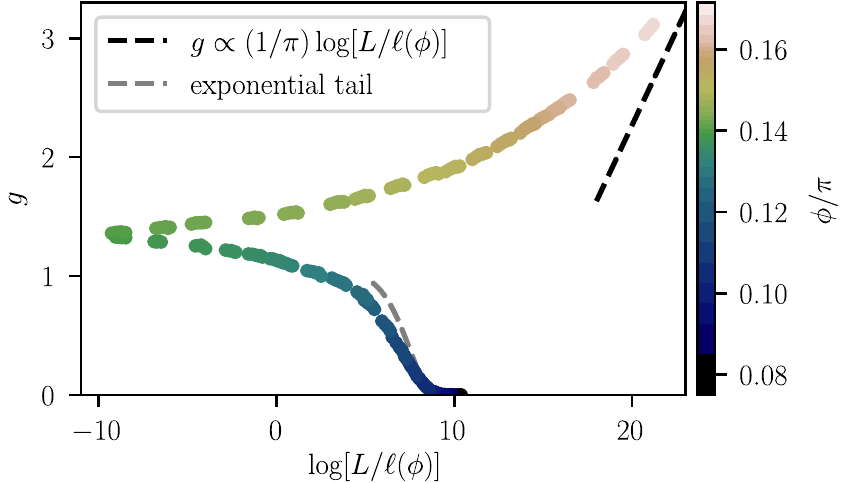}
 \vspace{-1.5em}
 \caption{Conductivity $g$ for the coherent-error network on a cylinder of length $L$ and circumference $M=5L$, averaged over $500$ to $10^5$ syndrome realizations.
 Error bars (2$\times$standard error) are imperceptible.
 The data following scaling curves $g[L/\ell(\phi)]$ shows that $\phi$ enters via a length scale $\ell(\phi)$~\cite{SM}.
 For the insulator, $g[L/\ell(\phi)]$  decays with $L$ and $g\propto e^{-2L/\ell(\phi)}$ for  $L\gg\ell(\phi)$.  
 For the metal, $g$ increases; the $g\propto \ln(L)/\pi$ class D asymptote~\cite{SenthilFisher2000,EversMirlin08}  is not yet reached for the accessible range of $L$. 
 The insulator-metal transition is at $\phi_\text{th}=(0.14\pm0.005)\pi$. 
 We observe $\mathcal{Q}=-1$ throughout the insulating phase. 
 }\label{fig:g}
\end{figure}

As $\phi$ increases, vortices proliferate. 
Generically, this gives a metal~\cite{Chalker01}, the phase we expect beyond $\phi_\text{th}$. 
(For the RBIM, $J$ being real precludes a metal~\cite{ReadLudwig2000}.)
To test this and find $\phi_\text{th}$, we study  $g$ for vortices drawn from $P_{s}$, sampling  using Ref.~\onlinecite{bravyi2018correcting}'s algorithm~\cite{SM}.
Our results are in Fig.~\ref{fig:g}. 
The $\mathcal{Q}=-1$ insulator persists up to $\phi_\text{th}=(0.14\pm0.005)\pi$, followed by a metal for $\phi>\phi_\text{th}$.
Both phases show single-parameter scaling: $\phi$ enters only via a length scale $\ell(\phi)$.
[For an insulator, $\ell(\phi)=\xi(\phi)$.]
While this qualitatively agrees with class D results~\cite{Medvedyeva10,Wang21}, for the metal $g[L/\ell(\phi)]$ increases slower towards  $\pi^{-1}\ln[L/\ell(\phi)]$ than predicted by the non-linear $\sigma$ model (the standard theory for the metallic phase~
\cite{SenthilFisher2000,ReadLudwig2000,EversMirlin08}). 
Establishing the insulator-metal phase diagram and $\phi_\text{th}$ are among our key results.
Conceptually, the network model phases offer coherent-error QEC phenomenology  akin to how RBIM phases do in the incoherent case.
Practically, since $dg/dL$, unlike $d\Delta/dL$ below, changes sign at $\phi_\text{th}$, 
the network model greatly facilitates identifying $\phi_\text{th}$.

\textit{From insulators to QEC:} 
We now establish $\phi_\text{th}$ as the coherent error threshold.
For this, we consider $\zeta_s=\frac{\mathcal{Z}_{1,s}}{\mathcal{Z}_{0,s}}$. 
In the Ising language, $\zeta_s$ is a disorder correlator~\cite{fradkin2017disorder,SM} since $\mathcal{Z}_{1,s}$ differs from $\mathcal{Z}_{0,s}$ by a row of sign-flipped bonds. 
We have $\zeta_s\propto e^{-\frac{1}{2}[E_{1,s}^{(0)}-E_{0,s}^{(0)}]L}$ for large $L$, with $E^{(0)}_{\hcl,s}$ the lowest energy of $\hat{\mathcal{H}}_{\hcl,s}$~\cite{SM}.
To evaluate $\zeta_s$, we consider the 1D free-fermion Hamiltonians that $\hat{\mathcal{H}}_{\hcl,s}$ gives for each $P$. 
These Hamiltonians have gap $\propto \xi^{-1}$ if the corresponding network is an insulator and their ground state has fermion parity $\nu\,\text{sgn}[\det(\mathcal{R}')]$ (with $\nu=\pm1$ set by $\mathcal{R}'$ conventions)~\cite{MerzChalker02,SM}. The latter fact not only allows one to view $\mathcal{Q}$ as their 1D topological invariant~\cite{kitaev2001unpaired,Fulga12,SM}, but, crucially, also implies that their number $n$ of excitations satisfies $(-1)^n=\nu\,\text{sgn}[\det(\mathcal{R}')]P$.

Since each flips a row of vertical bonds, $P$ and $\hcl$ effectively swap pbc and apbc for fermions. 
This is crucial when the network is a $\mathcal{Q}=-1$ insulator: from $\det(\Rtot')$ swapping sign, 
$\nu\sgn[\det(\Rtot')]=\chi_C(-1)^\hcl P$, with $\chi_C=\pm1$ set by $C_s$~\cite{SM}. 
Thus $(-1)^n=\chi_C(-1)^\hcl$, and $E_{1,s}^{(0)}-E_{0,s}^{(0)}\approx\chi_C/\xi$
up to $O(e^{-M/\xi})$ corrections from apbc vs. pbc energy differences. 
Hence, $\zeta_s\propto e^{-\chi_C L/2\xi}$, and $\Delta\propto e^{-z L/2\xi}$, with $z=2$ in the coherent and $z=1$ in the incoherent case (from $P_{\hcl,s}\propto |\mathcal{Z}_{\hcl,s}|^z$). 
The $\mathcal{Q}=-1$ insulator thus marks the error correcting phase. 
[For a $\mathcal{Q}=1$ insulator, $E_{1,s}^{(0)}-E_{0,s}^{(0)}=O(e^{-M/\xi})$: here QEC fails.]
This establishes $\phi_\text{th}$ as the coherent QEC threshold. 

\begin{figure}[t]
 \includegraphics[scale=1]{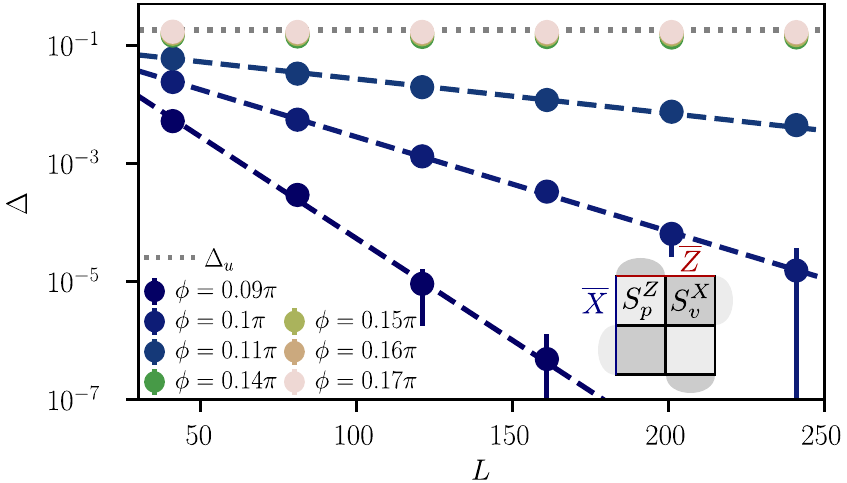}
 \vspace{-1.5em}
 \caption{Figure of merit $\Delta$ for the $L\times L$ planar geometry of recent experiments~\cite{krinner2022realizing,Google_SC} (cf. inset for $L=3,5$). 
 We averaged over $250$ to $2\!\times\!10^5$ syndrome realizations; error bars (2$\times$standard error) are imperceptible.
 $\Delta$ decays exponentially (dashed) with $L$ for $\phi<\phi_\text{th}$. 
 Above $\phi_\text{th}$, the data are consistent with $\Delta$ decaying as a power law to $\Delta_\infty(\phi)<\Delta_\text{u}=\frac{\pi-2}{2\pi}$. 
  }\label{fig:Delta}
\end{figure}

Fig.~\ref{fig:Delta} shows numerical results on $\Delta$ for the planar geometry of recent $L=3,5$ experiments~\cite{krinner2022realizing,Google_SC}. 
Our theory describing this system shows that our predictions hold beyond the cylinder~\cite{SM}: $\phi_\text{th}$ reflects bulk physics.

\textit{QEC in the metallic phase:} The metal for $\phi>\phi_\text{th}$, instead of the $\mathcal{Q}=1$ insulator, is a qualitatively new feature. 
While the $\mathcal{Q}=1$ insulator maps to a disordered phase~\cite{dennis2002topological,MerzChalker02}, the metal, if we generalize metallic disorder correlator results~\cite{ReadLudwig2000}, suggests a new QEC analog of quasi-long-range order where $\Delta$ decays nonexponentially with $L$.  
This is indeed what is seen in Fig.~\ref{fig:Delta}. 
The data are consistent with $\Delta=\lambda L^{-d_\Delta}+\Delta_\infty$, where $\lambda,d_\Delta,\Delta_\infty>0$ depend on $\phi$. 
Furthermore, we find $\Delta_\infty<\Delta_\text{u}=\frac{\pi-2}{2\pi}$, the value for uniform $\phi_s$ in the logical  $U_s=\exp(i\phi_s \Xlog)$ arising from QEC in this geometry~\cite{bravyi2018correcting,Florian20}.  

\textit{Conclusion:}
We have mapped surface codes with $\exp(i\phi X)$ [or $\exp(i\phi Z)$] coherent errors to a complex RBIM, and via its transfer matrix $\hat{\Mtot}_{\hcl,s}$, to quantum transport in 
a 2D Majorana network. 
The network yielded an insulator-metal phase diagram.
Linking the insulator's $\mathbb{Z}_2$-invariant to the topology of 1D fermions,  we explicitly mapped the insulator to the error-correcting phase and established the insulator-metal transition, at $\phi_\text{th}\approx 0.14\pi$, as the achievable storage threshold.
Such a high achievable threshold, with $\sin^2(\phi_\text{th})\approx 0.18>p_\text{th}\approx0.11$, explains why standard decoders, even if not optimal, can reach thresholds $\phi_\text{c}$ with $\sin^2(\phi_\text{c})\lesssim p_\text{th}$ as in Refs.~\cite{bravyi2018correcting,Florian20}.

The metal we found highlights fundamentally distinct coherent-error physics. It  maps to a new phase in QEC where, albeit as a power-law and to a nonzero $\Delta_\infty$ value, $\Delta$ decays with  $L$ (Fig.~\ref{fig:Delta}). 
This is markedly different from the incoherent case where, above threshold, $\Delta$ increases and saturates exponentially with $L$~\cite{Suchara14}.

\begin{acknowledgments}
This work was supported by EPSRC grant EP/V062654/1, a Leverhulme Early Career Fellowship, the Newton Trust of the University of Cambridge, and in part by the ERC Starting Grant No. 678795 TopInSy.
Our simulations used resources at the Cambridge Service for Data Driven Discovery operated by the University of Cambridge Research Computing Service (\href{www.csd3.cam.ac.uk}{www.csd3.cam.ac.uk}), provided by Dell EMC and Intel using EPSRC Tier-2 funding via grant EP/T022159/1, and STFC DiRAC funding  (\href{www.dirac.ac.uk}{www.dirac.ac.uk}).
\end{acknowledgments}


%%%%%%%%%%%%%%%%%%%%%%%%%%%%%%%%%%%%%%%%%%%%%%%%%%%%%%%%%%%%%%%%%%%%%%%

%
\pagebreak

\renewcommand{\thefigure}{S\arabic{figure}}
\renewcommand{\theequation}{S\arabic{equation}}

\onecolumngrid
\begin{center}
\large{\textbf{Supplemental Material for ``Coherent error threshold for surface codes from Majorana delocalization"}}
\end{center}
\vspace*{4em}
\twocolumngrid

\renewcommand\thefigure{S\arabic{figure}}
\setcounter{figure}{0}
\setcounter{equation}{0}

\section{Ising mapping}
\label{sec:stat-mech}

In the main text, we derive that $\braket{ \psi | C_s \Xlog^q U | \psi } = \mathcal{N} \mathcal{Z}_{q,s}$ with $\mathcal{N}=\prod_j\sqrt{i\sin\phi\cos\phi}$ and the Ising partition function $\mathcal{Z}_{q,s}$.
Here we provide further details on this.

We first expand $U = \prod_j (\openone \cos \phi + i X_j \sin \phi)$ in Pauli-$X$ strings.
The coefficient $A(\mathcal{X})$ of $X$-string $\mathcal{X}$ is $A(\mathcal{X})=(i \sin \phi )^{n(\mathcal{X})} (\cos \phi)^{N-n(\mathcal{X})}$, where $N$ is the number of qubits in the system and $n(\mathcal{X})$ is  the number of Pauli $X$ operators in $\mathcal{X}$.
Key to the mapping is to rewrite $A(\mathcal{X})$ via an indicator function $\eta_j(\mathcal{X})$ of the support $\text{supp}(\mathcal{X})$ (i.e., the sites on which $\mathcal{X}$ acts not as identity):
\begin{eqnarray}
A(\mathcal{X})&=&\prod_{j}\sqrt{i\cos\phi\sin\phi}\left(\sqrt{\frac{\cos\phi}{i\sin\phi}}\right)^{\eta_{j}(\mathcal{X})},\\ \eta_{j}(\mathcal{X})&=&\begin{cases}
-1 & j\in\text{supp}(\mathcal{X}),\\
1 & \text{otherwise.}
\end{cases}
\end{eqnarray}
In terms of $e^J=\sqrt{i\tan\phi}$ and $\mathcal{N}$, this is
\begin{equation}
A(\mathcal{X})=\mathcal{N} \prod_j e^{-J\eta_{j}(\mathcal{X})}.
\end{equation}
Since $\Xstab_v\ket{\psi}=\ket{\psi}$ and $\ket{\psi}$ is an eigenstate of $\Zlog$, the states $\ket{\psi_{\hcl,s}}=C_s \Xlog^\hcl\ket{\psi}$ form a complete orthonormal basis.
Hence, $\braket{ \psi | C_s \Xlog^q U | \psi }$ is the sum of coefficients $A(\mathcal{X})$ of terms from $U$ that  yield $\ket{\psi_{\hcl,s}}$.
Since $\ket{\psi_{\hcl,s}}= C_s \Xlog^\hcl \prod_{v\in V} \Xstab_v\ket{\psi}$ for any vertex set $V$, and this is the most general form in which $\ket{\psi_{\hcl,s}}$ can arise from $X$-strings,
\begin{multline}\label{Aqs_sum}
\braket{ \psi | C_s \Xlog^q U | \psi }=
\sum_{\{r_{v}=0,1\}}A(C_{s}\overline{X}^{q}\prod_{v}(\Xstab_v)^{r_{v}})\\=\mathcal{N}\sum_{\{r_{v}=0,1\}}\exp[-J\eta_j(C_{s}\overline{X}^{q}\prod_{v}(\Xstab_v)^{r_{v}})].
\end{multline}
From here, the Ising mapping follows via three observations. Firstly, by $X_j^2=\openone$, we have $\eta_{j}(\mathcal{X}\mathcal{X}')=\eta_{j}(\mathcal{X})\eta_{j}(\mathcal{X}')$ for any pair of $X$-strings $\mathcal{X}$ and $\mathcal{X}'$.
Hence in Eq.~\eqref{Aqs_sum} $\eta_j(C_{s}\overline{X}^{q}\prod_{v}(\Xstab_v)^{r_{v}})=\eta_j(C_{s}\overline{X}^{q})\eta_j(\prod_{v}(\Xstab_v)^{r_{v}})$.
Secondly, each qubit $j$ is uniquely specified by the adjacent nearest neighbor vertices $v$ and $v'$, and vice versa, which allows us to relabel $j\leftrightarrow v v'$.
Thirdly, since $\prod_{v\in V}\Xstab_v$ features $X_j$ if precisely one of the vertices adjacent to $j$ is in $V$ %
we find $\eta_{j}(\prod_{v}(\Xstab_v)^{r_{v}})=\sigma_{v}\sigma_{v^{\prime}}$ for $r_v\in\{0,1\}$, where $v$ and $v'$ on the right hand side are the nearest-neighbor pair specified by $j$ and where we defined $\sigma_{v}=-(-1)^{r_{v}}$.
This gives $\braket{ \psi | C_s \Xlog^q U | \psi } = \mathcal{N} \mathcal{Z}_{\hcl,s}$ with
\begin{equation}
 \mathcal{Z}_{\hcl,s}= \sum_{ \{ \sigma_v \} } e^{ - J \sum_{\langle v,v'\rangle} \eta_{vv'} \sigma_v \sigma_{v'} }
\end{equation}
as in the main text. Here $\langle v,v' \rangle$ labels nearest-neighbor vertices and $\eta_{vv'}=\eta_j(C_{s}\overline{X}^{q})$ (upon relabeling $j\leftrightarrow v v'$ and leaving the $s$ and $\hcl$ dependence implicit).
Note that $\mathcal{N}$ is independent of the syndrome $s$ and of $q$.

We evaluate $\mathcal{Z}_{q,s}$ using the transfer matrix $\hat{\Mtot}_{\hcl,s}$~\cite{sachdev_2011,fradkin2013field}.
This progresses through vertical slices of the $L\times M$ lattice (cf.~Fig.~\ref{fig:cyl_plan}) via operators $\hat{V}_{q,s}^{(k)}$ and $\hat{H}_{q,s}^{(k)}$ [respectively for the $k^\text{th}$ slice of vertical and horizontal bonds, cf.~Eq. (4) in the main text] in a $2^M$-dimensional transfer matrix Hilbert space (with basis states $\ket{\{\sigma_j\}}$) such that $\braket{\{\sigma_j\}| \hat{V}_{q,s}^{(k)}| \{\sigma'_{j}\} }=e^{-J\sum_j \eta_{\hcl,s}^{j,k}\sigma_j \sigma_{j+1}}\delta_{\{\sigma_j\},\{\sigma'_{j}\}}$ and $\braket{\{\sigma_j\}| \hat{H}_{q,s}^{(k)}| \{\sigma'_{j}\} }=e^{-J\sum_j \eta_{\hcl,s}^{j,k}\sigma_j \sigma'_j}$ (with sign $\eta_{\hcl,s}^{j,k}$ for vertical, resp., horizontal bond $j,k$).
This is achieved by $\hat{V}_{q,s}^{(k)} = \otimes_{j=1}^{M} \hat{v}_{\hcl,s}^{(j,k)}$ and $\hat{H}_{q,s}^{(k)} = \otimes_{j=1}^M A_{\hcl,s}^{(j,k)} \hat{h}_{\hcl,s}^{(j,k)}$ with $\hat{v}_{q,s}^{(j,k)} = \exp(-\kappa_{\hcl,s}^{j,k} Z_j Z_{j+1})$ and $\hat{h}_{\hcl,s}^{(j,k)} = \exp (-\tilde{\kappa}_{\hcl,s}^{j,k} X_j)$ [cf.~the main text for $A_{\hcl,s}^{(j,k)}$, $\kappa_{\hcl,s}^{j,k}$, and $\tilde{\kappa}_{\hcl,s}^{j,k}$].
The open boundary conditions at $x=0,L$ map to states $\ket{\alpha_{0,L}}$ (see Sec.~\ref{sec:BC} below) in terms of which $\mathcal{Z}_{\hcl,s} = \braket{ \alpha_L |\hat{\Mtot}_{\hcl,s}| \alpha_0 }$.
To expose the problem as free-fermionic, we Jordan-Wigner transform to introduce Majorana fermions $\hat{\gamma}_{2j-1} = (\prod_{k=1}^{j-1} X_k ) Z_j$, $\hat{\gamma}_{2j} = (\prod_{k=1}^{j-1} X_k) Y_j$~\cite{fradkin2013field}.
The matrices $\hat{v}_{\hcl,s}^{(j<M,k)}$ and $\hat{h}_{\hcl,s}^{(j,k)}$ are exponentials of Majorana bilinears (cf.~main text); not so for $\hat{v}_{\hcl,s}^{(M,k)} = \exp(i \kappa_{\hcl,s}^{M,k} \hat{P} \hat{\gamma}_{2M} \hat{\gamma}_1 )$ due to the parity operator $\hat{P}$ (a standard feature arising from pbc~\cite{fradkin2013field}). $\hat{P}$ may, however, be replaced by a sign when working in a fixed parity sector.

For operators $\hat{Q}=e^{\bm{\hat{\gamma}}^T q\bm{\hat{\gamma}}}$, we have $\hat{Q} \hat{\gamma}_j \hat{Q}^{-1}=\sum _m Q_{mn}\hat{\gamma}_m$ with the matrix $Q=e^{4q}$~\cite{kitaev2001unpaired}.
We can thus switch to the $2M\times 2M$ single-particle transfer matrix $\Mtot_{\hcl,s}$ since this captures $\hat{\Mtot}_{\hcl,s} \hat{\gamma}_j \hat{\Mtot}_{\hcl,s}^{-1} = \sum_l ( \Mtot_{\hcl,s} )_{lj} \hat{\gamma}_l$ in each of the parity sectors.
$\Mtot_{\hcl,s}$ is the product of slices $V_{\hcl,s}^{(k)}$ and $H_{\hcl,s}^{(k)}$ of single-particle transfer matrices, which in turn are direct sums $V_{\hcl,s}^{(k)} = \oplus_{j=1}^M v_{\hcl,s}^{(j,k)}$ and $H_{\hcl,s}^{(k)} = \oplus_{j=1}^M h_{\hcl,s}^{(j,k)}$ of
$2\times 2$ matrices $h_{q,s}^{(j,k)} = e^{2\tilde{\kappa}_{q,s}^{j,k} Y}$, $v_{q,s}^{(j<M,k)}=e^{2\kappa_{q,s}^{j,k} Y}$ and $v_{q,s}^{(M,k)}=e^{-2P \kappa_{q,s}^{M,k} Y}$.
Sec.~\ref{sec:BC} shows how to implement open or periodic transversal boundary conditions with $\ket{\alpha_{0,L}}$ being free-fermion (and thus fixed-parity) states  or simple combinations thereof; this justifies the use of single-particle transfer matrices.

\section{Boundary conditions}
\label{sec:BC}
\begin{figure}[b]
\includegraphics[width=\linewidth]{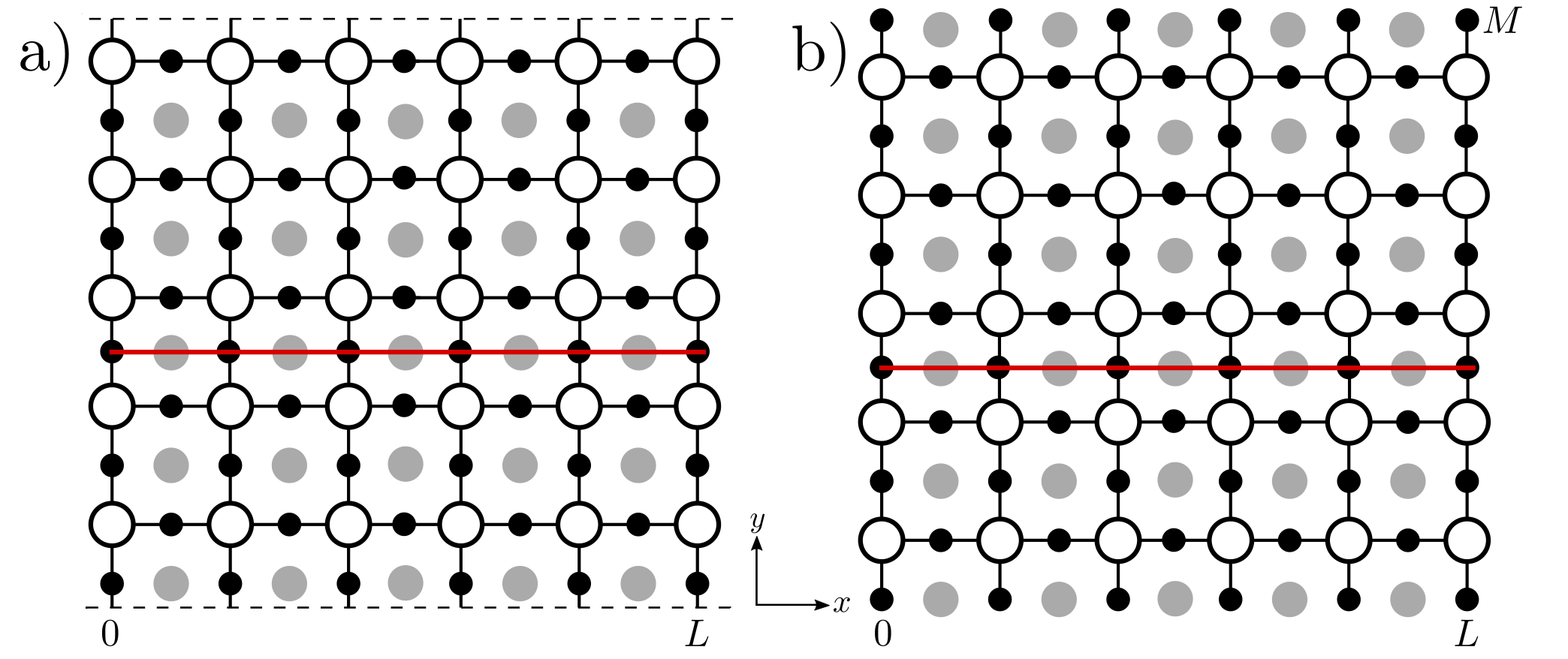}
\caption{Surface code on the cylinder (a, with upper edge continuing into the lower at the dashed line), and on the plane (b), each with boundaries parallel to the links (i.e., to the Ising bonds, cf.~Fig. 1 of the main text). As in the main text, black dots indicate qubits, white disks indicate $\Xstab_v$, gray disks show $\Zstab_w$, and the red line marks the path of $\Xlog$.
}
\label{fig:cyl_plan}
\end{figure}

The number of logical qubits in a surface code is set by the topology of the system, including its boundaries~\cite{kitaev2003fault,BravyiKitaev_SC,dennis2002topological,Fowler12}.
As in the main text, we focus on systems supporting a single logical qubit.
Here we provide details, in particular on how boundaries enter, for two such systems: a cylinder (Fig.~\ref{fig:cyl_plan}a) and a planar system (Fig.~\ref{fig:cyl_plan}b), each with boundaries, and shortest logical $\Xlog$ and $\Zlog$,  aligned with the bonds of the Ising model.
We will also comment on the planar system in Fig.~3 of the main text.

We start with the boundaries at $x=0,L$; these are the same for both systems in Fig.~\ref{fig:cyl_plan}.
These boundaries are chosen such that $\Xlog$ can terminate on them:
a string $\prod_j X_j$ connecting these boundaries violates no $\Zstab_w$.
The other feature of the $x=0,L$ boundaries is that, just as in the bulk, each qubit has two neighboring $\Xstab_v$.
Hence, the Ising model includes just the interaction terms $\sigma_v \sigma_{v'}$, specifically vertical bonds, at these boundaries.
Correspondingly, the first layer of the transfer matrix $\hat{\Mtot}_{\hcl,s}$ is $\hat{V}^{(1)}_{\hcl,s}=\otimes_{j=1}^M \hat{v}^{(j,1)}_{\hcl,s}$ where
$
\hat{v}_{\hcl,s}^{(j,1)}=e^{\kappa_{\hcl,s}^{j,1}Z_j Z_{j+1}}
$
has Pauli operators acting on the transfer matrix Hilbert space.
(Henceforth $Z_j$ and $X_j$ act on this space in this Section.)
Using this first layer, we must implement the sum over the first row of Ising spins, $\sum_{\sigma_v^{(1,1)}}\sum_{\sigma_v^{(2,1)}}\ldots \sum_{\sigma_v^{(M,1)}}$.
This is achieved by multiplying $\hat{V}^{(1)}_{\hcl,s}$ by $(|0_1\rangle+|1_1\rangle)\otimes(|0_2\rangle+|1_2\rangle)\otimes\ldots\otimes (|0_M\rangle+|1_M\rangle)=(\sqrt{2}|+\rangle)^{\otimes M}\equiv |\Psi_+\rangle$.
We proceed analogously at the $x=L$ boundary.
This gives
\begin{equation}
\mathcal{Z}_{\hcl,s}=\langle\Psi_+|\hat{\Mtot}_{\hcl,s}|\Psi_+\rangle.
\end{equation}

In the fermion language, $X_j=i\hat{\gamma}_{2j-1}\hat{\gamma}_{2j}$.
Hence, as the ground state of  $\hat{H}_X=-\sum_{j=1}^M X_j=-\sum_{j=1}^M i\hat{\gamma}_{2j-1}\hat{\gamma}_{2j}$, the state $|\Psi_+\rangle$  is a free-fermion state.

We next consider the horizontal ($y=0,M$) boundaries of the planar system in Fig.~\ref{fig:cyl_plan}b.
These boundaries have qubits with only one neighboring $\Xstab_v$.
This translates to a term $-J\sum_k [\eta^{(\hcl,s)}_{0,k}\sigma_v^{(1,k)}+\eta^{(\hcl,s)}_{M,k}\sigma_v^{(M,k)}]$ in the Ising Hamiltonian, i.e., to boundary magnetic field terms.
In the transfer matrix, this modifies the boundary gates for the first and $M$-th row of vertical bonds to
$
\hat{v}_{\hcl,s}^{(1,k)}=e^{\kappa_{\hcl,s}^{1,k}Z_1 Z_{2}-\kappa_{\hcl,s}^{0,k}Z_1}
$
and
$
\hat{v}_{\hcl,s}^{(M,k)}=e^{-\kappa_{\hcl,s}^{M,k}Z_M}.
$
Without the boundary magnetic fields, i.e., for $\kappa_{\hcl,s}^{0,k}=\kappa_{\hcl,s}^{M,k}=0$, all gates of the circuit commute with the parity $P=\prod_j X_j$.
The boundary magnetic fields break this symmetry because $Z_j$ is odd under $P$: $PZ_jP^\dagger =-Z_j$.

We now show that one can restore the $P$ symmetry of $\hat{\Mtot}_{\hcl,s}$ and map the problem back to the cylinder, at the expense of adding a zeroth auxiliary qubit in state $|1\rangle$ and extending $|\Psi_+\rangle$ into $|\Psi'_+\rangle=|\Psi_+\rangle\otimes|1\rangle$. (By $|1\rangle=(|+\rangle-|-\rangle)/\sqrt{2}$, the state $|\Psi'_+\rangle$ is a superposition of two free-fermion states.)
We take
$
\hat{v}_{\hcl,s}^{(1,k)}=e^{\kappa_{\hcl,s}^{1,k}Z_1 Z_{2}+\kappa_{\hcl,s}^{0,k}Z_0 Z_1}
$
and
$
\hat{v}_{\hcl,s}^{(M,k)}=e^{\kappa_{\hcl,s}^{M,k}Z_MZ_0}
$
and leave all other gates unaltered.
Denoting the thus extended transfer matrix by $\hat{\Mtot}_{\hcl,s}'$, we have $[Z_0,\hat{\Mtot}_{\hcl,s}']=0$, hence $\hat{\Mtot}_{\hcl,s}'|\Psi'_+\rangle=(\hat{\Mtot}_{\hcl,s}|\Psi_+\rangle)\otimes|1\rangle$ and thus $\mathcal{Z}_{\hcl,s}=\langle\Psi_+'|\hat{\Mtot}_{\hcl,s}'|\Psi_+'\rangle$ is the sum of two free-fermion overlaps (no cross terms arise due to $\hat{\Mtot}_{\hcl,s}'$ conserving the extended parity $P'=PX_0$).
From here the considerations about the cylinder apply.

Having discussed the geometries in Fig.~\ref{fig:cyl_plan}, we comment on the planar code in Fig. 3 of the main text.
As the codes in Fig.~\ref{fig:cyl_plan}, this also supports a single logical qubit but has $45^\circ$ rotated boundaries and shortest logical operators.
Our numerical results on $\Delta$ for this system show the error-correcting phase for $\phi<\phi_\text{th}$, with $\phi_\text{th}\approx0.14\pi$ matching (to the accuracy of our $\Delta$ simulations) the threshold from the Majorana network for cylinders as in Fig.~\ref{fig:cyl_plan}a.
This shows that the phases of QEC, and the threshold $\phi_\text{th}$, follow bulk physics and are largely insensitive to the details of boundaries.

\section{Network models from transfer matrices}
\label{sec:network_model}

In this section, we explain how describing the RBIM using transfer matrices results in the scattering network model that we use in the main text.
To this end, we first explain how pseudo-unitary single-particle transfer matrices give a current-conserving network model.
Then we show how this relates to the RBIM and the complex-coupling Ising model from the main text.

\begin{figure}
\includegraphics{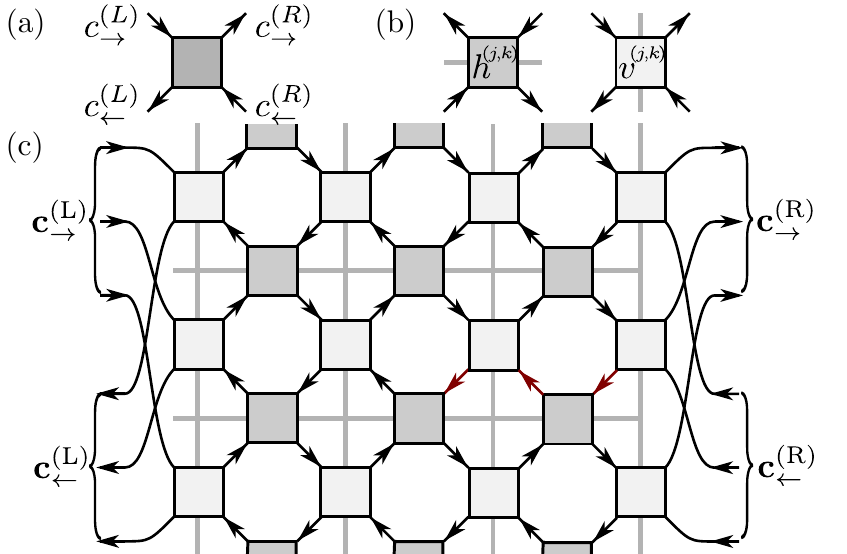}
\caption{(a) Transfer matrices relate modes on different sides of a system. The arrows indicate propagation direction.
(b) In the RBIM with real coupling $J$, both $h_{q,s}^{(j,k)}$ and $v_{q,s}^{(j,k)}$ are pseudounitary transfer matrices. (c) $h_{q,s}^{(j,k)}$ and $v_{q,s}^{(j,k)}$ thus result in a current-conserving network  model whose transfer matrix relates $\psi^{(\mathrm{L})} = (\mathbf{c}_\rightarrow^{(\mathrm{L})},\mathbf{c}_\leftarrow^{(\mathrm{L})} )^T$ and $\psi^{(\mathrm{R})} =( \mathbf{c}_\rightarrow^{(\mathrm{R})},\mathbf{c}_\leftarrow^{(\mathrm{R})} )^T$.
}
\label{fig:network_incoherent}
\end{figure}

The single-particle transfer matrix $\mathcal{M}$ takes a vector of amplitudes $\psi^{(\mathrm{L})} = ( \mathbf{c}_\rightarrow^{(\mathrm{L})}, \mathbf{c}_\leftarrow^{(\mathrm{L})} )^T$ to the vector $\psi^{(\mathrm{R})} =  (\mathbf{c}_\rightarrow^{(\mathrm{R})} , \mathbf{c}_\leftarrow^{(\mathrm{R})} )^T$ via $\psi^{(\mathrm{R})} = \mathcal{M} \psi^{(\mathrm{L})}$; here the transformation is from amplitudes $\psi^{(\mathrm{L})}$ on the left side of the system to $\psi^{(\mathrm{R})}$ on the right side, cf.~Fig.~\ref{fig:network_incoherent}.
We would like to view $\mathbf{c}_\rightarrow^{(\alpha)}$ and $\mathbf{c}_\leftarrow^{(\alpha)}$ as vectors of current amplitudes for modes propagating to the right and to the left, respectively, where the superscript $\alpha = \mathrm{L}, \mathrm{R}$ denotes the side of the system.
This view is useful if the corresponding current is conserved, i.e., if
\begin{equation}
 {\mathbf{c}_\rightarrow^{(\mathrm{R})}}^\dagger \mathbf{c}_\rightarrow^{(\mathrm{R})} - {\mathbf{c}_\leftarrow^{(\mathrm{R})}}^\dagger \mathbf{c}_\leftarrow^{(\mathrm{R})}
 = {\mathbf{c}_\rightarrow^{(\mathrm{L})}}^\dagger \mathbf{c}_\rightarrow^{(\mathrm{L})} - {\mathbf{c}_\leftarrow^{(\mathrm{L})}}^\dagger  \mathbf{c}_\leftarrow^{(\mathrm{L})},
\end{equation}
or equivalently if ${\psi^{(\mathrm{L})}}^\dagger (\mathcal{M}^\dagger Z \mathcal{M} - Z ) \psi^{(\mathrm{L})} = 0$.
For this to hold for any $\psi^{(\mathrm{\alpha})}$,  $\mathcal{M}$ must be pseudo-unitary: $\mathcal{M}^{-1} = Z \mathcal{M}^\dagger Z$.
Products of pseudo-unitary matrices retain this property, and network models arise from the multiplication of local pseudo-unitary transfer matrices (cf.~Fig.~\ref{fig:network_incoherent})~\cite{chalker1988percolation}.
The relation among $\mathbf{c}_\rightarrow^{(\alpha)}$ and $\mathbf{c}_\leftarrow^{(\alpha)}$ can be equivalently expressed via the scattering matrix $\mathcal{S}$~\cite{BeenRMT},
\begin{equation}
 \begin{pmatrix} \mathbf{c}_\leftarrow^{(\mathrm{L})} \\ \mathbf{c}_\rightarrow^{(\mathrm{R})} \end{pmatrix} =   \underbrace{ \begin{pmatrix} \mathcal{R} & \mathcal{T}' \\ \mathcal{T} & \mathcal{R}' \end{pmatrix} }_{\mathcal{S}} \begin{pmatrix} \mathbf{c}_\rightarrow^{(\mathrm{L})} \\ \mathbf{c}_\leftarrow^{(\mathrm{R})} \end{pmatrix} .
\end{equation}
This relates incoming to outgoing modes of the system; current conservation thus requires $\mathcal{S}^\dagger \mathcal{S} = \openone$.
The blocks of $\mathcal{S}$ are, respectively, matrices of transmission ($\mathcal{T}$ and $\mathcal{T}'$) and reflection ($\mathcal{R}$ and $\mathcal{R}'$) amplitudes.

The single-particle transfer matrix $\Mtot_{\hcl,s}$ for the Ising models we consider is built from $2\times 2$ matrices $h_{q,s}^{(j,k)} = e^{2\tilde{\kappa}_{q,s}^{j,k} Y}$, $v_{q,s}^{(j<M,k)}=e^{2\kappa_{q,s}^{j,k} Y}$ and $v_{q,s}^{(M,k)}=e^{-2P \kappa_{q,s}^{M,k} Y}$ [cf.~the main text and Sec.~\ref{sec:stat-mech}]. We illustrate them in Fig.~\ref{fig:network_incoherent}(b) for the 2D RBIM with real $J$, corresponding to incoherent errors in the surface code.
In this case, rotating $Y\to X$ gives real $h_{q,s}^{(j,k)}$ and $v_{q,s}^{(j,k)}$~\footnote{Although the $\tilde{\kappa}_{q,s}^{j,k}$ are complex for $\eta_{vv'}^{(q,s)}=-1$, the resulting $h_{q,s}^{(j,k)}$ remains real.} which are pseudo-unitary.
They hence result in a current-conserving network model~\cite{MerzChalker02}, shown in Fig.~\ref{fig:network_incoherent}(c).

As shown in the main text, coherent errors imply a complex $J$ in the Ising model, for which
$v_{q,s}^{(j,k)}$ and $h_{q,s}^{(j,k)}$ are not pseudo-unitary.
Still, they yield a current-conserving network: $h_{q,s}^{(j,k)}$ is unitary, hence is a scattering matrix conserving horizontal one-way current.
Similarly, $v_{q,s}^{(j,k)}$ satisfies $Z {v_{q,s}^{(j,k)}}^{-1} Z = -{v_{q,s}^{(j,k)}}^\dagger$, hence $v_{q,s}^{(j,k)} X$ is pseudo-unitary. Hence, $v_{q,s}^{(j,k)}$ describes a transfer matrix with swapped propagation directions on one side, i.e., it describes one-way vertical propagation.
We show the resulting network model in Fig. 1 of the main text. %

\section{2D insulators and gapped 1D systems}
\label{app:2Dto1D}

In the main text, we invoked links between the 2D insulating networks and 1D gapped Hamiltonians.
Here we discuss these relations, using an approach that bridges between the results of Ref.~\onlinecite{Fulga12} linking 1D and 2D topological phases via scattering matrices (albeit with 1D Hamiltonians of different origin than ours) and Ref.~\onlinecite{MerzChalker02} which provides early topological insights linking 2D networks and 1D systems in symmetry class D.

As in the main text, we work with a system on a cylinder (Fig.~\ref{fig:cyl_plan}a).
For simplicity, here we consider that we have a local free-fermion system to begin with, unlike in our main text where  fermions originate from a Jordan-Wigner transformation.
As a result, the transfer matrix $\hat{\Mtot}$ and Hamiltonian $\hat{\mathcal{H}}$  now do not have explicit parity dependence.
In particular, now $\hat{\mathcal{H}}$, not only its restriction to a parity sector, is quadratic in the $\hat{\gamma}_j$.
[We consider this scenario only to simplify presenting the key 1D to 2D relations. The explicit parity dependence of $\hat{\mathcal{H}}_{\hcl,s}$ was important when we arrived to $(-1)^n=\chi_C (-1)^\hcl$ in the main text.]
We take $\hat{\Mtot}$ and $\hat{\mathcal{H}}$ to be disordered, but omit the subscript $\hcl,s$ for brevity.

We start by describing some well-known facts about 1D free-fermion Hamiltonians~\cite{kitaev2001unpaired}.
Being quadratic in the $\hat{\gamma}_j$ we write $\hat{\mathcal{H}}=\frac{i}{2}\bm{\hat{\gamma}}^T A\bm{\hat{\gamma}}$ in terms of the vector $\bm{\hat{\gamma}}=(\hat{\gamma}_1,\hat{\gamma}_2,\ldots,\hat{\gamma}_{2M})$ and a real antisymmetric matrix $A$.
One may write $iA=W^T \bm{E}W$ where $\bm{E}=-\bm{\varepsilon}\otimes Y$ with $\bm{\varepsilon}=\text{diag}(\varepsilon_1,\varepsilon_2,\ldots,\varepsilon_M)$ a diagonal matrix of $0\leq\varepsilon_j\leq\varepsilon_{j+1}$ excitation energies, $Y$ the second Pauli  matrix, and $W$ a $2M\times2M$ real orthogonal matrix.

Using this, we can show that $\hat{\mathcal{H}}$ is gapped whenever the network is an insulator.
Recall, $\hat{\mathcal{H}}$ is defined by $\hat{\Mtot}\hat{\Mtot}^\dagger=e^{-L\hat{\mathcal{H}}}$.
Using the correspondence between an operator $\hat{Q}=e^{\bm{\hat{\gamma}}^T q\bm{\hat{\gamma}}}$ and matrix $Q=e^{4q}$ in the
linear transformation $\hat{Q} \hat{\gamma}_j \hat{Q}^{-1}=\sum _m Q_{mn}\hat{\gamma}_m$~\cite{kitaev2001unpaired}, the matrix $\Mtot$ for $\hat{\Mtot}$ satisfies
$
\Mtot\Mtot^\dagger=\exp[-2iLA].
$
This implies
\begin{equation}\label{eq:Mhyp1}
\Mtot\Mtot^\dagger=W^T
\begin{pmatrix}
\cosh(2L\bm{\varepsilon}) & -i \sinh(2L\bm{\varepsilon})\\
i \sinh(2L\bm{\varepsilon}) & \cosh(2L\bm{\varepsilon})
\end{pmatrix}
W.
\end{equation}
Since $\Mtot$ is pseudounitary, so is $\Mtot\Mtot^\dagger$, which implies that $W$ commutes with $\openone_M\otimes Z$, hence $W=\text{diag}(W_1,W_2)$.
A similar decomposition yields $\Mtot^\dagger \Mtot={W'}^T \exp (2 L \bm{\varepsilon} \otimes Y) W'$, where we used that $\Mtot^\dagger \Mtot$ and $\Mtot\Mtot^\dagger$ share the same eigenvalues.
We can readily read off the polar decomposition~\cite{BeenRMT} $\Mtot = W^T \exp(L \bm{\varepsilon} \otimes Y) W'$ and the corresponding transmission matrix $\mathcal{T} = W_1^T \sech (L\bm{\varepsilon}) W_1'$, and, using $g = (L/M) \Tr [ \mathcal{T}^\dagger \mathcal{T} ]$, the average conductivity
\begin{equation}
g=\frac{L}{M}\left\langle\sum_{j=1}^M\frac{1}{\cosh^2(L{\varepsilon}_j)}\right\rangle.
\end{equation}
In an insulator $g\propto e^{-2L/\xi}$ upon increasing system size at fixed aspect ratio $L/M$; here $\xi$ is the localization length.
This requires the smallest energy ${\varepsilon}_1$ to satisfy $\lim_{M\rightarrow\infty} {\varepsilon}_1>0$.
(${\varepsilon}_1$ becomes increasingly non-random upon increasing the system size~\cite{BeenRMT,MerzChalker02}.)
Hence, $\hat{\mathcal{H}}$ has a gap $\varepsilon_\text{gap}=\lim_{M\rightarrow\infty} {\varepsilon}_1 \propto \xi^{-1}$ (average and typical $\varepsilon_1$ may differ by a factor of order unity~\cite{EversMirlin08}).
This gap does not close unless the network  delocalizes ($\xi\rightarrow \infty$).

Next we link the ground-state fermion parity of the 1D gapped $\hat{\mathcal{H}}$ to the scattering matrix of the 2D insulator.
For a given boundary condition, the fermion parity of the ground state is $P_\text{GS}=\sgn(\text{Pf}A)=\det(W)$ where Pf is the Pfaffian~\cite{kitaev2001unpaired}.
[In the main text we also use that this implies $(-1)^n=\det(W)P$ for the number $n$ of excitations: $n=0$ in the ground state and changing $n$ by one flips $P$.]
In our case, from  $W=\text{diag}(W_1,W_2)$ %
we have $P_\text{GS}=\det(W_1W_2)$.
Comparing the polar decompositions of the scattering and of the transfer matrix~\cite{BeenRMT}, one finds that, in a gauge where $\Rtot'$ is real, $\sgn[\det(\Rtot')]=\nu \det(W_1W_2)$ with $\nu=\pm1$ set by conventions used for $\Rtot'$.
[For example, changing the sign conventions for one of the outgoing modes re-gauges the mapping between $W_j$ and $\Rtot'$, taking $\nu\rightarrow-\nu$.]
Hence, $P_\text{GS}=\nu\sgn[\det(\Rtot')]$.

This result also allows us to equate the topological invariants of the 1D ground state~\cite{kitaev2001unpaired} and of the 2D scattering network~\cite{MerzChalker02,Fulga12}.
(The 1D topological interpretation is further supported by the 2D insulator implying the entanglement area-law for the 1D ground state~\cite{Jian22}.)
In 1D, the topological invariant $\mathcal{I}$ compares $P_\text{GS}$ for periodic and antiperiodic boundary conditions (pbc and apbc, respectively)~\cite{kitaev2001unpaired}.
Concretely, $\mathcal{I}=P_\text{GS}^\text{pbc}P_\text{GS}^\text{apbc}$ with $\mathcal{I}=-1$ in a topological phase~\cite{kitaev2001unpaired}.
(The topological phase supports Majorana endmodes  for open boundary conditions.)
From above, $\mathcal{I}=\sgn[\det(\Rtot'_\text{pbc}\Rtot'_\text{apbc})]$.

In 2D, we first discuss symmetry class DIII of time-reversal invariant Majorana systems.
The reflection matrix $\Rtot^{\prime\text{DIII}}$ is real and antisymmetric~\cite{Fulga12}.
The $\mathbb{Z}_2$ invariant is $\mathcal{Q}=\sgn[\text{Pf}(\Rtot^{\prime\text{DIII}}_\text{pbc})\text{Pf}(\Rtot^{\prime\text{DIII}}_\text{apbc})]$~\cite{Fulga12,FulgaDIII}.
Our system is in class D: we have a Majorana system without time-reversal symmetry.
From this, we get a class DIII system by considering it together with its time-reversed copy.
The corresponding doubled reflection matrix can be written as
$\Rtot^{\prime\text{DIII}}=\left(\begin{smallmatrix}
0 & \Rtot' \\
\Rtot'^T & 0
\end{smallmatrix}\right)$
in terms of the reflection matrix $\Rtot'$ of our class D system.
Since $\Rtot'$ is an $M\times M$ matrix, now $\text{Pf}(\Rtot^{\prime\text{DIII}})=(-1)^{M(M-1)/2}\det(\Rtot')$.
Hence, the topological invariant reads
$\mathcal{Q}=\sgn[\det(\Rtot'_\text{pbc}\Rtot'_\text{apbc})]$.
This establishes $\mathcal{I}=\mathcal{Q}$, the link between 1D and 2D topological invariants.
Since $\mathcal{I}$ cannot change without gap closing, and $\varepsilon_\text{gap}\propto\xi^{-1}$, we also find that  $\mathcal{Q}$ cannot change without delocalization.

\section{Syndrome sampling}

To sample from $P_s$, we use Bravyi \textit{et al.}'s fermion linear optics (FLO) algorithm~\cite{bravyi2018correcting,Florian20}.
(We could also Monte Carlo sample via our fermionic quantum circuit, however this method suffers from typical Monte-Carlo issues: a long burn-in time and correlations between realizations.)
We summarize the essential ideas; for a detailed derivation and discussion of subtleties we refer to Refs.~\onlinecite{bravyi2018correcting,Florian20}.

\begin{figure}
\includegraphics[width=\columnwidth]{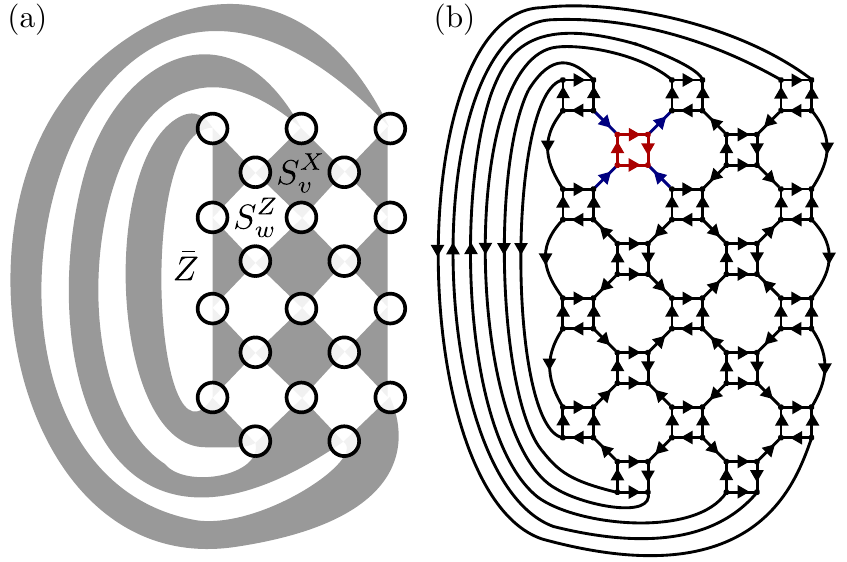}
\caption{(a) An embedding of a surface code from a cylinder into the plane. White circles mark the qubits, grey faces mark the $S_v^X$ stabilizers, and white faces mark the $S_w^Z$ stabilizers and the logical $\bar{Z}$ operator. (b) A corresponding Majorana graph. Marked in red is the $C4$ encoding of a qubit and in blue the edges
that connect this qubit to the other qubits. \label{fig:flat_cylinder}}
\end{figure}

Since we apply only $X$ rotations, all stabilizers $S_v^X$ remain in the $+1$ state.
It is thus sufficient to sample from the measurement outcomes of the $\Zstab_w$ which we can obtain via the outcomes $m_j=\pm1$ of measuring $Z_j$ for each physical qubit and classically computing the corresponding outcomes for the $\Zstab_w=\prod_{j\in w} Z_j$. %
We sample the $Z_j$ measurement outcomes using conditional probabilities: Given a string $m_1 ,\dots , m_{j-1}$ of previous outcomes from qubits $1,\ldots j-1$, outcome $m_j$ has conditional probability
\begin{equation}
 p_j (m_j|m_1 ,\dots m_{j-1} ) = \frac{p_j (m_1,\dots,m_j)}{p_{j-1} (m_1,\dots,m_{j-1})} ,
 \label{eq:conditional_probability}
\end{equation}
which can be computed using FLO~\cite{bravyi2018correcting}.
To this end, the qubits are encoded in the $C4$ encoding: each qubit $j$ is encoded into four Majorana modes $\hat{\gamma}_{4j-3}, \dots, \hat{\gamma}_{4j}$ and stabilized with the qubit stabilizer $\bar{S}_j = - \hat{\gamma}_{4j-3} \hat{\gamma}_{4j-2} \hat{\gamma}_{4j-1} \hat{\gamma}_{4j}$.
The Majorana modes of all qubits are then placed in a Kastelyn oriented graph following Ref.~\onlinecite{Florian20}.
Since this requires embedding the surface code into a plane, if working on a cylinder (the geometry we use to compute $g$ in the network model) we need to unwrap the system  around its base as illustrated in Fig.~\ref{fig:flat_cylinder}(a).
We show the corresponding Majorana graph in Fig.~\ref{fig:flat_cylinder}(b).
The graph has the following property: if a fermionic state $\ket{\psi}$ is initialized according to the orientation of the arrows in between different qubits, which means that  an arrow from $\hat{\gamma}_a$ to $\hat{\gamma}_b$ indicates that $i \hat{\gamma}_a \hat{\gamma}_b \ket{\psi} = \ket{\psi}$, and then projected to the $+1$ space of all qubit stabilizers $\bar{S}_j$, the resulting state is the $C4$ encoding of the logical $\ket{0}_L$ state.

To utilize this graph for sampling syndromes, i.e., computing the conditional probability~\eqref{eq:conditional_probability}, we use that a state initialized according the the graph Fig.~\ref{fig:flat_cylinder}(b), but \emph{before} projecting onto the qubit stabilizers $\bar{S}_j$, is a fermion Gaussian state.
Projecting the state to any $\bar{S}_j = 1$ space would destroy that property since $\bar{S}_j$ is a four-Majorana operator.
Instead, we use that the $C4$ encoding of the coherent error rotation commutes with $\bar{S}_j$ and the state remains a fermion Gaussian state after the coherent rotation; in the simulation we thus apply the rotations first.
The final trick is to exploit that for an individual qubit, even though we leave the space of fermion Gaussian states when projecting to $\bar{S}_j = 1$, we immediately return back to it after projectively measuring $Z_j$ for  qubit $j$ since in the $C4$ encoding, both $Z_j$ and $\bar{S}_j Z_j$ are two-Majorana operators.
This way we can progress through the code qubit-by-qubit (while maintaining graph connectivity~\cite{bravyi2018correcting,Florian20}),  projecting to $\bar{S}_j = 1$ followed by the $Z_j$ measurement and thus maintaining a fermion Gaussian state, and sampling $m_j$ according to \eqref{eq:conditional_probability}. ($m_1$ can be sampled tossing a fair coin~\cite{bravyi2018correcting}.) %
From each $\{m_j\}$ realization we can compute a realization of the $S_w^Z$ outcomes, and thus sample the syndrome $s$.
In practice, we use this via defining $\eta_j(C_s)=m_j$ to get the Ising bond signs for a $C_s$ consistent with the sampled syndrome $s$.
\section{From 1D fermions to error correction via disorder correlators}
\label{App:1DtoEC}

A key result in the main text is that in the error correcting phase, the free-fermion Hamiltonians that $\hat{\mathcal{H}}_{\hcl,s}$ gives for each $P$ are gapped and  topologically nontrivial.
Here we provide further details on this.
We also expand on the link between $\Delta$ and disorder correlators and show that the typical disorder correlator and $\Delta$ behave similarly.

The energies $E_{\hcl,s}^{(n)}$ and eigenvectors $\ket{\varphi_n}\!$ of the 1D Hamiltonian $\hat{\mathcal{H}}_{\hcl,s}$\! in $\hat{\Mtot}_{\hcl,s}\hat{\Mtot}_{\hcl,s}^\dagger=e^{-L\hat{\mathcal{H}}_{\hcl,s}}$ enter the singular value decomposition of $\hat{\Mtot}_{\hcl,s}$: we have
\begin{equation}\label{eq:SVD}
\hat{\Mtot}_{\hcl,s}=\sum_n e^{-E_{\hcl,s}^{(n)}L/2}\ket{\varphi_n}\!\bra{\widetilde{\varphi}_n},
\end{equation}
with $\ket{\varphi_n}\!$ being the left singular vectors.
(The $\bra{\widetilde{\varphi}_n}$ are the right singular vectors, i.e., the eigenvectors of $\hat{\Mtot}_{\hcl,s}^\dagger\hat{\Mtot}_{\hcl,s}$.)
Eq.~\eqref{eq:SVD} shows that, for long-time evolutions [$L\varepsilon^{(1)}_{\hcl,s}\gg1$ with $\varepsilon^{(1)}_{\hcl,s}$ the smallest excitation energy of $\hat{\mathcal{H}}_{\hcl,s}$],
a generic initial state is projected to the ground state $\ket{\varphi_0}$\! of $\hat{\mathcal{H}}_{\hcl,s}$\! by the quantum circuit $\hat{\Mtot}_{\hcl,s}$.
(When there is a degenerate ground space, a projection to this takes place instead.)
In particular, from Eq.~\eqref{eq:SVD} follows that  $\zeta_s=\frac{\mathcal{Z}_{1,s}}{\mathcal{Z}_{0,s}}$ satisfies $\zeta_s\propto e^{-\frac{1}{2}[E_{1,s}^{(0)}-E_{0,s}^{(0)}]L}$ for large $L$, with $E_{\hcl,s}^{(0)}$ the lowest energy of $\hat{\mathcal{H}}_{\hcl,s}$ and prefactor set by $x=0,L$ boundary conditions (cf.~Sec.~\ref{sec:BC}).

As in the main text, we take a system on a cylinder (Fig.~\ref{fig:cyl_plan}), and consider $\hat{\mathcal{H}}_{\hcl,s}$ for each parity $P=\pm1$, where it is free-fermionic.
For each $P=\pm1$, we write $\hat{\mathcal{H}}^{(P)}_{\hcl,s}=\frac{i}{2}\bm{\hat{\gamma}}^T A_{\hcl,s}^{(P)}\bm{\hat{\gamma}}$ with a real antisymmetric matrix $A^{(P)}_{\hcl,s}$\!.

Were $iA^{(P)}_{\hcl,s}$ a full single-particle Hamiltonian (instead of applying for a given parity $P$), we would have $\sgn\text{Pf}A^{(P)}_{\hcl,s}=\det W^{(P)}_{\hcl,s}$ for the fermion parity in the ground state~(cf.~Sec.~\ref{app:2Dto1D}).
The ground state would then satisfy $P=\det[W^{(P)}_{\hcl,s}]$, which generalizes for $n$ excitations to $(-1)^n =\det[W^{(P)}_{\hcl,s}]P$ (also cf.~Sec.~\ref{app:2Dto1D}).
This, together with $\hat{\mathcal{H}}^{(P)}_{\hcl,s}$ applying for a given $P$, shows that the ground state of $\hat{\mathcal{H}}^{(P)}_{\hcl,s}$ is a valid state only when $\det[ W^{(P)}_{\hcl,s}]P=1$; otherwise only states with an odd number of excitations of $\hat{\mathcal{H}}^{(P)}_{\hcl,s}$ are valid in that parity sector.
While this might seem unusual, it is a standard feature when, as $\hat{\mathcal{H}}_{\hcl,s}$, a fermion Hamiltonian arises via Jordan-Wigner transformation from a spin system with pbc~\cite{fradkin2013field}.

As we next explain, this feature is at the core of how, for gapped $\hat{\mathcal{H}}^{(P)}_{\hcl,s}$, the error correcting phase and the phase where error correction fails are distinguished by the topological invariant $\mathcal{I}=\det[(W_{\hcl,s}^{(P)})_{\text{pbc}}(W_{\hcl,s}^{(P)})_{\text{apbc}}]$ (cf.~Sec.~\ref{app:2Dto1D}).
The key observation is that $\hat{P}$ enters $\hat{\mathcal{H}}_{\hcl,s}$ via $\hat{v}_{\hcl,s}^{(M,k)}=e^{i\hat{P}\kappa_{\hcl,s}^{M,k,}\hat{\gamma}_{2M}\hat{\gamma}_{1}}$: it sets the sign of a row of transversal bonds along the cylinder.
Hence, flipping the value of $P$ switches between pbc and apbc for $\hat{\mathcal{H}}^{(P)}_{\hcl,s}$.

For $\mathcal{I}=-1$, this implies $\det W^{(P)}_{\hcl,s}=-\det W^{(-P)}_{\hcl,s}$ and hence $\det[W^{(P)}_{\hcl,s}]P$ has the same sign for either of $P=\pm1$: the ground state of $\hat{\mathcal{H}}^{(P)}_{\hcl,s}$ is either a valid state for both $P=\pm1$ or for neither.
The other key observation is that flipping $\hcl$ swaps these two possibilities:  as $\Xlog$ runs along the cylinder, flipping $\hcl$  also flips the signs along a row of transversal bonds, $\det W^{(P)}_{\bar{\hcl},s}=-\det W^{(P)}_{\hcl,s}$.
Without knowing $C_s$ we cannot tell which sign of $\det[W^{(P)}_{\hcl,s}]P$ goes with which value of $\hcl$, since for $C_s$ and $C_s'=C_s\Xlog$ the conclusion would be the opposite.
Hence, we can write only $(-1)^n=\det [W^{(P)}_{\hcl,s}]P=\chi_C(-1)^\hcl$, with $\chi_C=\pm 1$, depending on $C_s$.
[The $C_s$ dependence is not a shortcoming: the purpose of $\Delta$, Eq.~(1) of the main text, is merely to tell whether $\hcl=\pm1$ are distinguishable.]
For $\chi_C=1$, the ground state ($n=0$) corresponds to $q=0$; for $q=1$ the lowest-energy state has $n=1$ excitations. For $\chi_C=-1$ it is vice versa. Hence,
\mbox{$E^{(0)}_{1,s}-E^{(0)}_{0,s}=\chi_C\varepsilon_\text{gap}$,}
where $\varepsilon_\text{gap}$ is the gap of $\hat{\mathcal{H}}^{(P)}_{\hcl,s}$~\footnote{The behavior of the system generalizes that familiar from simple limits of the transverse field Ising chain with pbc, $H=-\sum_j J_j Z_j Z_{j+1}-h\sum_j X_j$ with $J_{j\neq 1}=J$, $J_1=J(-1)^q$. For $\mathcal{I}=-1$ the simple limit is $h=0$ with $J>0$; there for $q=0$ both fully polarized states are ground states with all bonds satisfied, yielding energy $E_\text{GS}$. Conversely, $q=1$ flips a bond thus any lowest-energy state has an unsatisfied bond and energy $E_\text{GS}+\varepsilon_\text{gap}$. For $\mathcal{I}=1$, the limit is $J=0$ and $h\neq 0$. Now the ground state is unique and $H$ has no $q$ dependence.}. ($\varepsilon_\text{gap}$ is $P$-independent, up to corrections of order $e^{-\varepsilon_\text{gap}M}$ from pbc vs. apbc energy differences.)
In particular, for a gapped, topologically nontrivial $\hat{\mathcal{H}}^{(P)}_{\hcl,s}$, we have from $\zeta_s\propto e^{-\frac{1}{2}\chi_C\varepsilon_\text{gap}L}$ that $\Delta\propto e^{-\frac{z}{2} \varepsilon_\text{gap}L}$ with $z=2$ for the coherent and $z=1$ for the incoherent case (from $P_{\hcl,s}\propto |\mathcal{Z}_{\hcl,s}|^z$).

For a gapped topologically trivial ($\mathcal{I}=1$) $\hat{\mathcal{H}}^{(P)}_{\hcl,s}$,
$\det W^{(P)}_{\hcl,s}$ has the same sign for pbc and apbc.
Hence, $\det[W^{(P)}_{\hcl,s}]P$ depends on $P$ but not on $\hcl$.
For one $P$ value,  $E^{(0)}_{q,s}$ is the ground state energy $E_\text{GS}$, for the other value it is  $E_\text{GS}+\varepsilon_\text{gap}$; in either case $E^{(0)}_{1,s}-E^{(0)}_{0,s}=0$ up to corrections of order $e^{-\varepsilon_\text{gap}M}$ (from $\hcl$ swapping between pbc and apbc). Hence, $\Delta$ tends to a constant with $L$.

We thus found that a topologically trivial gapped $\hat{\mathcal{H}}^{(P)}_{\hcl,s}$ corresponds to the phase where error correction fails, while a topologically nontrivial gapped $\hat{\mathcal{H}}^{(P)}_{\hcl,s}$ marks the error correcting phase.

\vspace*{-1em}
\subsection{\textit{$\Delta$ and disorder correlators}}
\vspace*{-1em}

In  $\zeta_s=\frac{\mathcal{Z}_{1,s}}{\mathcal{Z}_{0,s}}$, the numerator is obtained from the denominator by flipping a row of bonds along the cylinder.
It is thus a correlation function $\langle\mu(x_1)\mu(x_2)\rangle_\beta$ of ``disorder operators" $\mu(x)$ evaluated at $x_1=0$, $x_2=L$~\cite{fradkin2017disorder}.
[Here $\langle\ldots\rangle_\beta$ denotes (a formal, in the coherent case) thermal average.]
The choice ${\mathcal{Z}_{0,s}}$ in the denominator and ${\mathcal{Z}_{1,s}}$ in the numerator is arbitrary: the underlying $C_s$ ($\hcl=0$) is no more valid reference Pauli string than $C_s\Xlog$ ($\hcl=1$).
One may thus consider the more generic $\zeta_{\hcl,s}=\frac{\mathcal{Z}_{{\bar\hcl},s}}{\mathcal{Z}_{\hcl,s}}$ instead.
[Here ${\bar \hcl}=\hcl+1\text{ (mod $2$)}$.]

In the incoherent case, we have $P_{\bar \hcl,s}/P_{\hcl,s}=\zeta_{\hcl,s}$; in the coherent case $P_{\bar \hcl,s}/P_{\hcl,s}=|\zeta_{\hcl,s}|^2$, thus in each case the probability ratio is, or is closely linked to, a disorder correlator.
In what follows, for brevity, we simply call $C^{(\hcl,s)}_{\mu\mu}\equiv P_{\bar \hcl,s}/P_{\hcl,s}$ disorder correlator in both cases.

Intuitively, $C^{(\hcl,s)}_{\mu\mu}$ measures the free energy cost of inserting a row of flipped bonds in the system, and is thus expected to decay exponentially in an ordered phase where vortices, i.e., flipped $\Zstab_w$, are rare, and be $L$-independent in the disordered phase where vortices proliferate.
This intuitive picture suggest using disorder correlators to detect the error correcting phase~\cite{dennis2002topological}.

This intuition tacitly considers the reference configuration [the denominator in $C^{(\hcl,s)}_{\mu\mu}$] to represent the typical case, i.e., the one with probability $\max_\hcl P_{\hcl,s}$.
In this case, $\frac{P_{\bar \hcl,s}}{P_{\hcl,s}}\propto  e^{-z L/2\xi}$ (with $z=2$ for the coherent and $z=1$ the incoherent case): this is the intuitive exponential decay.

However, when performing a syndrome (i.e., vortex disorder) average, one considers  all inequivalent bond configurations consistent with each $s$.
For the syndrome averaged disorder correlator, this gives
\begin{equation}
C^\text{(avg)}_{\mu\mu}=\sum_{s,\hcl}P_{\hcl,s} C^{(\hcl,s)}_{\mu\mu}=\sum_{s,\hcl}P_{\hcl,s}\frac{P_{\bar\hcl,s}}{P_{\hcl,s}}=1.
\end{equation}
This result was established in Ref.~\onlinecite{MerzChalkerDC} for the RBIM on the Nishimori line, i.e., for the incoherent case, using a different reasoning.
It shows that when sampling Pauli strings [i.e., $\eta^{(\hcl,s)}_{vv'}$] according to $P_{\hcl,s}$ (in the incoherent case this is equivalent to sampling $P_s$ by generating $X_j$ with probability $p$), rare events, with probability $\min_\hcl P_{\hcl,s}$, occur where $C^{(\hcl,s)}_{\mu\mu}$ increases exponentially, and these events precisely balance out the much more frequent exponential decay, occurring with probability $\max_\hcl P_{\hcl,s}$.

To capture the behavior in this more frequent case, one can use the ``typical" disorder correlator $C^\text{(typ)}_{\mu\mu}$\!, with
\begin{equation}
\ln C^\text{(typ)}_{\mu\mu}\equiv\sum_{s,\hcl}P_{\hcl,s} \ln C^{(\hcl,s)}_{\mu\mu}=\sum_{s,\hcl}P_{\hcl,s}\ln\frac{P_{\bar\hcl,s}}{P_{\hcl,s}}.
\end{equation}
The logarithm in $\ln C^{(\hcl,s)}_{\mu\mu}$ renders the exponentially increasing and decreasing cases of $C^{(\hcl,s)}_{\mu\mu}$ into factors of similar magnitude, hence the sum is dominated by the typical (of probability $\max_\hcl P_{\hcl,s}$) terms.
Mathematically, $\ln C^\text{(typ)}_{\mu\mu}$ has the appealing feature of being (the minus of) a symmetrized Kullback-Leibler divergence, a measure of the difference between the distributions $P_{\hcl,s}$ and $P_{\bar\hcl,s}$.
These features make $C^\text{(typ)}_{\mu\mu}$ a suitable measure to detect the error correcting phase~\cite{dennis2002topological}:
in this phase,  $C^\text{(typ)}_{\mu\mu}$ decays exponentially with $L$, while it becomes $L$-independent in the disordered phase.

\begin{figure}
 \includegraphics[scale=1]{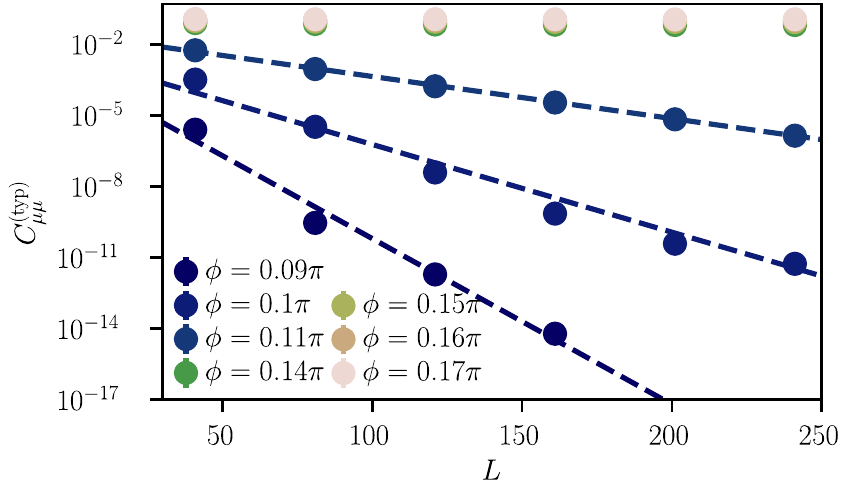}
 \vspace*{-0.3cm}
 \caption{Typical disorder correlator $C^\text{(typ)}_{\mu\mu}$ for the $L\times L$ planar geometry in Fig. 3 of the main text.
 We averaged over $250$ to $2\!\times\!10^5$ syndrome realizations; error bars show 2$\times$standard error.
 The behavior of is similar to $\Delta$: exponential decay (dashed) with $L$ for $\phi<\phi_\text{th}$, and, above $\phi_\text{th}$, trend consistent with a power law decay to a $\phi$-dependent value.
  }\label{fig:Cmumu}
\end{figure}

The figure of merit $\Delta$ in Eq.\,(1) of the main text is closely linked to $C^\text{(typ)}_{\mu\mu}$: here, instead of using the logarithm to suppress the contribution from the terms with $\min_\hcl P_{\hcl,s}$ probability, we simply discard them: we can write $\Delta=\sum_{s} \min_\hcl P_{\hcl,s}$ as
\begin{equation}
\Delta = \sum_{s} \max_\hcl P_{\hcl,s}\frac{\min_\hcl P_{\hcl,s}}{\max_\hcl P_{\hcl,s}}=\sum_{s} \max_\hcl P_{\hcl,s} \min_\hcl\frac{ P_{\bar\hcl,s}}{P_{\hcl,s}}.
\end{equation}
The behavior of $\Delta$ and $C^\text{(typ)}_{\mu\mu}$ is indeed similar as can be seen by comparing Fig.~3 of the main text and Fig.~\ref{fig:Cmumu}.

\vspace*{-1em}
\subsection{Decoherence of logical errors}
\vspace*{-1em}

A key question for coherent errors is the decoherence of logical noise~\cite{Beale18,bravyi2018correcting,iverson2020coherence,Florian20}: does the post-QEC state $\ket{\psi^{\prime}}$ approach the result of logical Pauli operations on the initial state $\ket{\psi}$ as $L\to\infty$?
Equivalently, does the post-syndrome-measurement state $\ket{\psi_s}= \pi_sU\ket{\psi}/\sqrt{P_s}$ (with projector $\pi_s$ for syndrome $s$) approach either of $C_s\Xlog^\hcl\ket{\psi}$ ($\hcl=0,1$), i.e., the result of $\ket{\psi}$ having suffered a Pauli error consistent with $s$?
By $|\braket{C_s\Xlog^\hcl\psi|\psi_s}|^2=P_{\hcl,s}/P_s$, we find that $\Delta$ measures the syndrome average of the corresponding (in)fidelity.
When $\Delta\to 0$ for $L\to\infty$, as is the case for $\phi<\phi_\text{th}$, we find that the logical noise decoheres with $L$. For $\phi>\phi_\text{th}$, the result $\Delta=\lambda L^{-d_\Delta}+\Delta_\infty$ (where $\lambda,d_\Delta,\Delta_\infty>0$ and depend on $\phi$) implies a power-law decoherence of logical noise, but with a residual level of coherence (set by $\Delta_\infty$) remaining even as $L\to\infty$.

\vspace*{-1em}
\section{Raw conductivity data}
\label{App:graw}
\vspace*{-1em}

\begin{figure}
 \includegraphics[scale=1]{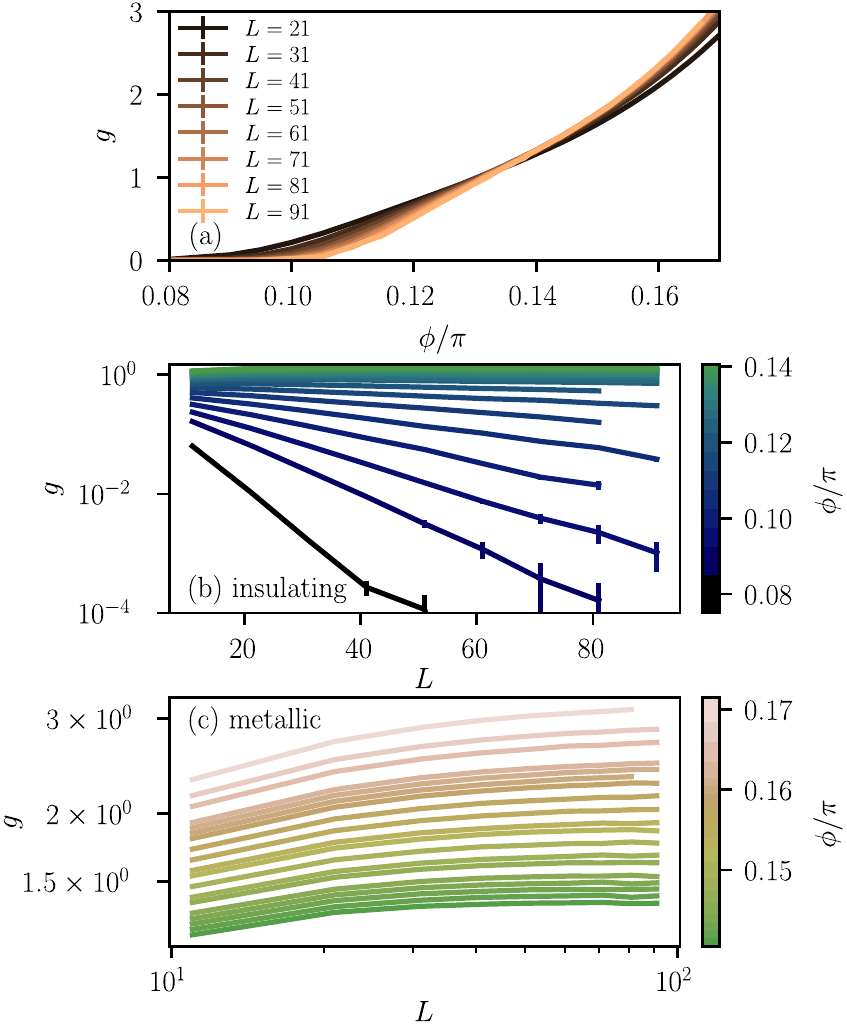}
 \vspace*{-0.3cm}
 \caption{Conductivity $g$ for the coherent-error network. As in the main text, the data are for a cylinder of length $L$ and circumference $M=5L$, averaged over $500$ to $10^5$ syndrome realizations.
 Error bars show 2$\times$standard error.
  }\label{fig:graw1}
\end{figure}

\begin{figure}
 \includegraphics[scale=1]{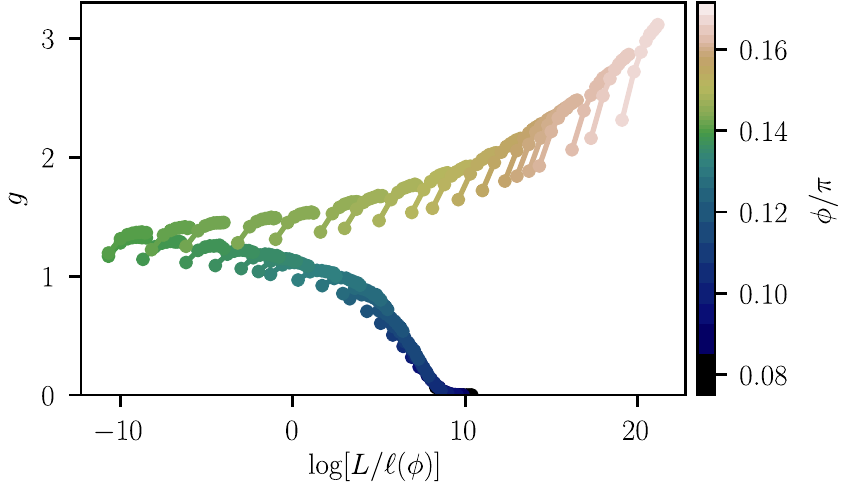}
 \vspace*{-0.3cm}
 \caption{Conductivity scaling curve with ballistic data included. The settings are the same as in Fig.~\ref{fig:graw1}.
  }\label{fig:graw2}
\end{figure}

\afterpage{\clearpage}
In Fig.~2 of the main text we show the scaling curves $g[L/\ell(\phi)]$ for the insulating and metallic phases.
Fig.~\ref{fig:graw1} shows the corresponding raw (i.e., unscaled) data.
Fig.~\ref{fig:graw1}a shows $g$ versus $\phi$; the curves cross at $\phi=\phi_\text{th}$.
Figs.~\ref{fig:graw1}b,c show $g$ versus $L$ in the insulating and metallic phases, respectively.
The conductivity is ballistic ($g\propto L$) for small $L$; the decaying (for the insulator) or diffusive ($g\propto \ln L$, for the metal) $g$ sets in only for $L$ beyond the localization length or mean free path, respectively.
Fig.~2 of the main text includes data only in this regime.
Fig.~\ref{fig:graw2} shows $g[L/\ell(\phi)]$ with the small-$L$ ballistic $g$ included.


\begin{thebibliography}{60}
\bibitem [{cyl()}]{cylinderfn}%
  \BibitemOpen
  \href@noop {} {}\bibinfo {note} {For boundary conditions allowing
  $\overline{X}$ along the cylinder, the $\!\ket{\alpha_r}$ are also
  free-fermion states~\cite{SM}.}\BibitemShut {Stop}%
\bibitem [{pi4()}]{pi4fn}%
  \BibitemOpen
  \href@noop {} {}\bibinfo {note} {This is due to the QEC problem being
  equivalent (upon $X_j\mapsto-X_j$) for $\phi$ and $-\phi$, and being related
  by a $\hcl,s$-independent factor $\prod_jX_j$ for $\phi$ and
  $\phi+\pi/2$.}\BibitemShut {Stop}%
\end{thebibliography}
\end{document}